\documentclass[a4paper,10pt]{article}
\pdfoutput=1

\usepackage{jcappub}

\usepackage[style=base, singlelinecheck=off, font=small, labelfont=bf,
justification=raggedright]{caption}
\usepackage{subfigure}

\usepackage[utf8]{inputenc}

\usepackage{float}
\usepackage{bm}
\usepackage{latexsym}
\usepackage{graphicx}
\usepackage[export]{adjustbox}
\usepackage{color}
\usepackage[dvipsnames]{xcolor}
\usepackage{verbatim}
\usepackage{array}
\usepackage{todonotes}
\usepackage{appendix}

\newcommand{\xpol}{\ensuremath{\tt Xpol}}
\newcommand{\commask}{{\tt {GAL70}}}
\newcommand{\commaskk}{{\tt {PL76}}}
\newcommand{\pysm}{\ensuremath{\tt PySM}}
\newcommand{\Nside}{\ensuremath{\tt Nside}}
\newcommand{\planck}{{\it Planck\/}}

\newcommand{\healpix}{\ensuremath{\tt HEALPix}}
\newcommand{\nested}{\ensuremath{\tt NESTED}}
\def\GHz{\ifmmode $\,GHz$\else \,GHz\fi}
\def\mKs{\ifmmode $\,mK\,s$^{1/2}\else \,mK\,s$^{1/2}$\fi}
\def\muKs{\ifmmode \,\mu$K\,s$^{1/2}\else \,$\mu$K\,s$^{1/2}$\fi}

\title{\ensuremath{\tt Deep\,\, Needlet}: A CNN based full sky component separation method in Needlet space}
\author[1,2,3,4]{Debabrata Adak}
\emailAdd{adak@iac.es}
\affiliation [1]{Instituto de Astrof\'{\i}sica de Canarias, E-38200 La Laguna, Tenerife, Spain}
\affiliation[2]{Departamento de Astrofísica, Universidad de La Laguna, E-38206 La Laguna, Tenerife, Spain}
\affiliation[3]{The Institute of Mathematical Sciences, CIT Campus, Chennai 600113, India}
\affiliation[4]{Homi Bhabha National Institute, Training School Complex, Anushakti Nagar, Mumbai 400085, India}
\date{\today}
\abstract{
One of the key steps in Cosmic Microwave Background (CMB) data analysis is component separation to recover the CMB signal from multi-frequency observations contaminated by foreground emissions.  Needlet Internal Linear Combination (NILC) is one of the successful methods that applies the minimum variance estimation technique to a set of needlet-filtered frequency maps to recover CMB. 
In this work, we develop a deep convolutional neural network (CNN) model to recover CMB temperature map from needlet-filtered frequency maps over the full sky. The network operates on a multi-resolution representation of spherical data, capturing localised features in both pixel and harmonic space, and is designed to preserve the rotational invariance of the CMB signal.
 The network model is trained on realistic simulations at \planck\ frequencies, which include CMB temperature maps generated using cosmological parameters sampled within a 2$\sigma$ standard deviation around the \planck\ best-fit values. We demonstrate the network performance for simulations that exhibit different foreground complexities. The recovered CMB temperature map closely follows the true signal with some residual leakage near the Galactic plane. The TT power spectrum is accurately reconstructed up to multipoles of approximately $\ell\sim 1100$. A minor residual systematics remain at smaller scales. Compared to the NILC method, the network shows reduced residual foreground contamination in the recovered CMB map.
 Once validated on the simulations, the network is applied to \planck\ PR3 intensity data. The resulting CMB map is consistent with the CMB maps from the \planck\ legacy products, including those produced using the NILC and SMICA pipelines. 
This work demonstrates a powerful component separation method to clean spherical signal data from multi-resolution wavelet-filtered maps.

}

\begin{document}

\maketitle

\section{Introduction}\label{Introduction}
The Cosmic Microwave Background (CMB) anisotropy is the snapshot of the Universe after the epoch of recombination. The map of temperature anisotropy of CMB, in addition to CMB polarisation data, contains a ton of information about the early and late Universe, which is proven to be crucial in the study of cosmology and useful to build the concordance model of the Universe. Therefore, ground-based, balloon-borne and satellite missions are being designed, planned and operated to observe the sky at millimeter (mm)  and sub-millimeter (submm) bands \citep{ Polarbear:2014,QUIJOTE:2015,Galitzki:2018,gualtieri2018spider,bicep3:2018,ACT:2020, SPT:2020,planck-I:2018,PICO:2019, Hazumi:2019, Adak:2022}. However, the CMB observations are contaminated by Galactic and extragalactic emissions of thermal dust, synchrotron, anomalous microwave emission (AME), free-free, CO, Cosmic Infrared Background (CIB) and  Sunyaev–Zeldovich (SZ) effects (see \cite{Ichiki:2014} for review). Therefore, component separation methods play a critical role in recovering CMB maps from foreground-contaminated data. Over the last decades, many component separation algorithms have been developed to clean the CMB using multi-frequency observations \citep{Tegmark:1997, Delabrouille:2003, Eriksen:2008, J.Delabrouille:2009, Basak:2011, Fernandez-Cobos:2012, SILC:2016, Remazeilles:2020, adak:2021, Carones:2024}. Some of these algorithms are blind that do not assume any particular model of foreground emissions \citep{Tegmark:1997, Delabrouille:2003}. Some methods are parametric \citep{Eriksen:2008, Stompor:2008} and some works at interface of both \citep{Remazeilles:2020,adak:2021,Carones:2024}. The main motivation for developing different algorithms is to reduce residual leakage of foregrounds and instrument noise in the recovered CMB map.  

The Needlet Internal Linear combination (NILC) method is one of the successful methods extensively applied to \planck\ and WMAP data analysis \cite{Basak:2013}. It is designed to improve the performance of Internal Linear combination (ILC, \cite{Tegmark:1997}). The method exploits the fact that foregrounds follow different spectra other than CMB and are uncorrelated with CMB anisotropy. The Galactic foregrounds are dominant at large angular scales. On the other hand, the instrument noise and extragalactic components are crucial contaminants at small angular scales. The needlet is a type of wavelet that enjoys localisation in both the pixel domain and harmonic space. Therefore minimum variance technique on needlet-filtered maps performs better than pixel/harmonic space  ILC methods \citep{Tegmark:1997, kim:2008}. However, the uncertainties in CMB calibration, loss of information at large scales for cut-sky data and any chance correlation between CMB and other components introduce bias in this method \citep{J.Delabrouille:2009, Dick:2010}. 

Foregrounds are non-Gaussian and anisotropic in nature. Therefore, the minimum variance estimation technique is perhaps non-optimal in the presence of higher-order statistics introduced by foregrounds. Therefore, an artificial neural network can be helpful since the method can approximate arbitrary functions with trainable hyperparameters and non-linear transformations \cite{russel:2010}. Deep convolutional neural networks (CNN) have demonstrated outstanding performance in image processing and classification problems. An extensive application of the technique in different machine learning (ML) models is seen in cosmological data analysis, including CMB \citep{2010A&A...520A..87N,2017arXiv171102033R, Thorne:2021, 2019MNRAS.490.1055C, 2019A&C....2800307C, 10.1093/mnras/stac736, 2020MNRAS.494.5761H, 2024arXiv240405794N, 2021PhRvD.104d3529G, 2021MNRAS.508.4600C, Jeffrey:2022, 2024arXiv240403557M, 2024arXiv240418897G, 2024arXiv240418100P,Yan:2023bjq,Lambaga:2025nbw}. The cosmological application of neural networks demands that the network architectures should not encode either direction or position according to the cosmological principle. In particular, for application of CNN for CMB data analysis, CNN should be efficient to maintain rotational invariance on $\mathbb{S}^2$ \cite{2008MNRAS.383..539M}.  
Therefore, several CNN techniques are developed to perform convolution over spherical data \cite{2018arXiv180110130C,2019arXiv190102039M,2019A&A...628A.129K, 2022arXiv220610385Y, deepsphere_cosmo}. In particular, a full sky component separation algorithm is developed in \cite{Petroff:2020} using DeepSphere algorithm \cite{deepsphere_cosmo}. The method is found to be sub-optimal at angular scales of $\ell > 900$.  More recently \cite{Wang:2022} applies the U-Net  model \cite{Ronneberger:2015} approximating the \healpix\ maps on $\mathbb{S}^2$ to $\mathbb{R}^2$ plane of size 4\,\Nside$\times$3\,\Nside. The method shows good agreement between the recovered power spectrum and the ground truth at $\ell < 900$. In this work, we develop a network model to apply to needlet-filtered frequency data at mm and submm bands to perform the component separation. We aim to investigate the potential improvements achieved by utilising multi-resolution needlet-filtered maps. Instead of providing the full Fourier information of the data in a single set of training images, in this work, we feed the CNN with a set of multiple band-filtered images. The goal of this approach is to enable the network to learn non-Gaussian foreground features across multiple angular scales, thereby improving the accuracy and robustness of foreground subtraction. 

We develop a U-Net architecture, a feedforward neural network model that has been inspired by neuroscience image classification in \cite{Ronneberger:2015} and previous works on component separation \citep{Petroff:2020, Wang:2022, Yan:2023bjq}. The details of the network architecture are discussed in Section~\ref{cnn_arc}. 
We adopt the concept of Wang et al. \cite{Wang:2022} to approximate \healpix\ maps to 2D plane of size 4\,\Nside$\times$3\,\Nside. This facilitates the application of CNN in $\mathbb{R}^2$ while approximately preserving the rotational invariance of the original spherical data.

The paper is organised as follows. We discuss the needlet decomposition of the maps and NILC minimum variance estimation technique in Section~\ref{nilc}. In Section~\ref{cnn_arc} we discuss the CNN model, namely \ensuremath{\tt Deep\,\, Needlet} developed in this work. The foreground models used in network training and testing and \planck\ data are described in Section~\ref{data}. The data used in the input and output layers of the model is preprocessed for easy application of CNN, which is described in Section~\ref{preprocessing}. Section~\ref{reults} discusses the training procedure of the network and test results on different foreground complexities and frequency configurations. The application of the network model to \planck\ data and corresponding results are discussed in Section~\ref{planck_application}. Finally, we summarise the main findings of this work in Section~\ref{summary}. In Appendix~\ref{sec:App1}, we discuss the bias in the recovered CMB map in more detail.

\section{Needlet ILC method}\label{nilc}
In this section, we discuss the methodology of CMB recovery using NILC pipeline.
CMB observations provide frequency maps convolved with the beam at respective channels. In harmonic space, the observed frequency maps can be written as,

\begin{equation}\label{eq:freq_maps}
    X^{Obs}_{\ell m,\nu} = a_{\nu}b_{\ell, \nu}X^{CMB}_{\ell m} + b_{\ell, \nu}X^{FG}_{\ell m} + X^{Noise}_{\ell m,\nu}, 
\end{equation}
 were $ a_{\nu}$ is the CMB calibration coefficient and  $b_{\ell,  \nu}$ is the beam transfer function at frequency $\nu$. In thermodynamic unit $ a_{\nu} = 1$ up to the calibration uncertainty. The $X^{CMB}_{\ell m}, X^{FG}_{\ell m}$ and $X^{Noise}_{\ell m}$ are the harmonic coefficients of CMB, total foregrounds and instrument noise, respectively. 

\begin{figure}
\includegraphics[width=\linewidth]{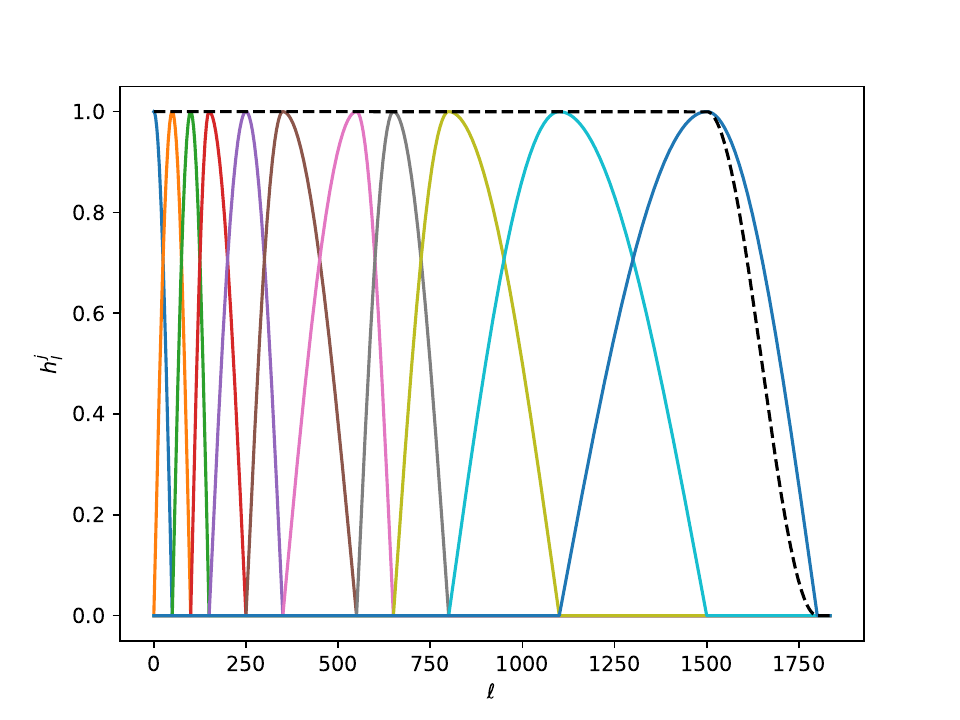}
\caption{The band shapes of needlet filters designed in Table~\ref{tab:needle_specifications}. The black dashed line displays normalisation of the needlet bands, i.e., the total filter applied in needlet decomposition of the map.}
\label{fig:needlet_bands}
\end{figure}

The harmonic coefficients, $X^{Obs}_{\ell m,\nu}$ of the frequency maps are first deconvolved with their native beam FWHM by dividing their respective beam transfer function $b_{\ell, \nu}$ and convolved with a common beam by multiplying the corresponding beam transfer function $b^{c}_{\ell}$ as,
\begin{equation}
    X_{\ell m, \nu} =  \frac{b^{c}_{\ell}}{b_{\ell, \nu}} X^{Obs}_{\ell m,\nu} .
\end{equation}\label{eq:deconv_conv}
We choose a common beam FWHM = 7.27$^{'}$ which sets the beam resolution of output maps. The resulting harmonic coefficients, $X_{\ell m, \nu}$ undergo bandpass filtering using needlet filters, $ h^{j}_{\ell}$ which decompose the frequency maps into a set of filtered maps in harmonic space:  $X^j_{\ell m, \nu} \longmapsto X_{\ell m, \nu}h^{j}_{\ell}$. This procedure corresponds to filtering the frequency maps using a spherical needlet defined in terms of $h^{j}_{\ell}$ as, 
\begin{equation}
    \psi_{jk}(\hat{n}) = \sqrt{\lambda_{jk}} \sum_{\ell = 0}^{\ell_{max}}\sum_{m = -\ell}^{\ell}h^{j}_{\ell}Y^{*}_{\ell m}(\hat{n})Y_{\ell m}(\hat{\xi}_{jk}),
\end{equation}
The $\lambda_{jk}$s are the cubature weights and $\hat{\xi}_{jk}$ denotes the cubature points on the sphere corresponding to the choice of \Nside\ suitable for $j$th needlet band. The needlet filters $h^{j}_{\ell}$ are localised in the harmonic space that follows the relation,
\begin{equation}
    \sum_{j} (h^{j}_{\ell})^2 = 1.
\end{equation}
This ensures that the power in the data after forward and inverse needlet transformations remains preserved.
In this project, we design the needlet filters as follows: 
\begin{equation}\label{eq:needlet_filter}
    \centering
    h^j_{\ell} = \left\{ \begin{tabular}{cl}
    $\cos\left[ \left(\frac{\ell^{j}_{peak}- \ell}{\ell^{j}_{peak}- \ell_{min}^{j}}\right)\frac{\pi}{2}\right]$ & \text{for } $\ell_{min}^{j}$ $\leq$ $\ell$ $< \ell_{peak}^{j}$ \\
    &\\
    $1$ & \text{for } $\ell =$ $\ell_{peak}^{j}$\\
    &\\
    $\cos\left[ \left(\frac{\ell-\ell^{j}_{peak} }{\ell^{j}_{max}- \ell_{peak}^{j}}\right)\frac{\pi}{2}\right]$ & \text{for } $\ell_{peak}^{j}<$ $\ell$ $\leq$ $\ell_{max}^{j}$. 
    \end{tabular} \right.
\end{equation}
The $\ell^{j}_{max}, \ell^{j}_{min}$, $\ell^{j}_{peak}$ for different bands are listed in Table~\ref{tab:needle_specifications}. The band shapes of filter functions are shown in Figure~\ref{fig:needlet_bands}.

Finally, the needlet coefficient maps are obtained from the filtered harmonic coefficients, $X^j_{\ell m, \nu}$ as,
\begin{equation}\label{eq:needlet_coeff_maps}
    \beta_{j,\nu} (\hat{n}) = \sum_{\ell, m} X^j_{\ell m, \nu}Y_{\ell m}(\hat{n}).
\end{equation}
The needlet coefficients are computed on \healpix\ grids having a suitable resolution parameter \Nside\ equal to the smallest power of 2 larger than $l^{j}_{max}$/2 that are listed in the last column of Table~\ref{tab:needle_specifications}.
\begin{table}
\caption{List of needlet bands used in the analysis.}  \centering \begin{tabular}{c c c c c} \hline\hline
  Band index & $l_{min}$ & $l_{peak}$ & $l_{max}$ & \Nside\ \\ [5ex]
  \hline 
  1 & 0   & 0 & 50 & 32\\ 
  2 & 0   & 50 & 100 & 64 \\ 
  3 & 50 & 100 &150 & 128 \\ 
  4 & 100 & 150 & 250 & 128\\
  5 & 150 & 250 & 350 & 256\\
  6 & 250 & 350 & 550 & 512 \\ 
  7 & 350 & 550 & 650 & 512 \\ 
  8 & 550 & 650 & 800 & 512 \\ 
  9 & 650 & 800 & 1100 & 1024 \\
  10 & 800 & 1100 & 1500 & 1024 \\
  11 & 1100 & 1500 & 1800 & 1024 \\
  [1ex] \hline
\end{tabular} 
\label{tab:needle_specifications} 
\end{table}

At each needlet band, all frequency maps are linearly combined by weights $w^{j}_{\nu}$ to get minimum variance estimated CMB needlet coefficient maps,
\begin{equation}
    \hat{\beta}^{NILC}_{j} (\hat{n}) = \sum_{\nu} w^{j}_{\nu} (\hat{n})\beta_{j, \nu}(\hat{n}),
\end{equation}
where the weights at each pixel are computed from the band covariance matrix, $C^{j}_{\nu,\nu^{'}} (\hat{n}) = <\beta_{j,\nu}(\hat{n})\beta_{j,\nu^{'}}(\hat{n})W_{j}^{PIX}(\hat{n})>$ as, 
\begin{equation}
    w^{j}_{\nu} (\hat{n}) = \frac{\sum_{\nu} [C^{j}_{\nu,\nu^{'}}(\hat{n})]^{-1}}{\sum_{\nu,\nu^{'}} [C^{j}_{\nu,\nu^{'}}(\hat{n})]^{-1}}.
\end{equation}
The window function $W_{j}^{PIX}(\hat{n})$ defines the domain over which the covariance is computed and this is a Gaussian smoothing function in pixel domain having different beam width at different band depending on independent modes at corresponding needlet band \citep{McCarthy:2023hpa}. 
The final NILC CMB map is then reconstructed by inverse filtering $\hat{\beta}^{NILC}_{j} (\hat{n})$ with same filters with $h^{j}_{\ell}$ used in decomposition of frequency maps following,
\begin{equation}\label{EQ:NEEDLET-TO-CMB}
    \hat{X}^{CMB} (\hat{n}) = \sum_{\ell m} \sum_{j} \big(\hat{\beta}^{NILC}_{j} (\hat{n}) h^{j}_{\ell}\big) Y^{*}_{\ell m} (\hat{n}).
\end{equation}
The final recovered map  $\hat{X}^{CMB} (\hat{n})$ contains true CMB along with some contamination from foreground and noise residuals. 
\section{CNN architecture}\label{cnn_arc}
The fundamental building block of a feedforward neural network is a model to map a set of images $\textbf{X} = \{X_1, X_2,...X_{n}\}$ to final output image $f_{w,b}(\textbf{X}) \simeq Y$ with a set of trainable hyperparameters $\{w,b\}$ connected to multiple layers. The information propagates forward layer to layer from input images to output images with connected neurons. In this particular problem, the inputs are needlet coefficients of frequency maps $\beta_{j,\nu} (\hat{n})$, and the output is needlet coefficients of the CMB map. We use 143 \GHz\ as the frequency channel for the network’s output. We adopt the U-Net architecture \citep{Ronneberger:2015} for application to spherical data represented in needlet space, referring to this machine learning model as \ensuremath{\tt Deep\,\, Needlet}. Details on the input and output data structures, designed to work with \healpix\ maps while preserving rotational invariance, are discussed in Section~\ref{preprocessing}. 
\begin{table}
\caption{The U-Net architecture used in this work for input and output maps of \Nside = 1024, corresponding to the highest-resolution needlet-filtered maps. Scaled versions of the same architecture are applied to maps at other \Nside\ maps, matching the resolution of the respective needlet bands. The output dimensions of the feature maps at each hidden layer are provided in the third and fifth columns of the table. The third value in each output shape denotes the number of convolutional filters used at that layer. The \Nside\ values corresponding to the input and output maps are listed in the leftmost column.}  
\begin{tabular}{|p{1.2cm}| p{2.8cm}|p{3cm}|p{3.4cm}|p{2.9cm}|}
\hline
\multicolumn{1}{|c|}{\Nside}&\multicolumn{2}{|c|}{Encoder block} & \multicolumn{2}{|c|}{Decoder block}\\
\cline{2-5}
&Layers& Output layer shape &Layers& Output layer shape\\
\hline
1024&Conv2D block ($c_1$) & (2048, 1536, 8)& DeConv2D block ($d_9$)& (4096, 3072, 1) \\
512&Conv2D block ($c_2$) & (1024, 768, 16)& DeConv2D block ($d_8$)& (2048, 1536, 8) \\
256&Conv2D block ($c_3$) & (512, 384, 32) &DeConv2D block ($d_7$)& (1024, 768, 16) \\
128&Conv2D block ($c_4$)& (256, 192, 64)& DeConv2D block ($d_6$)& (512, 384, 32) \\
64&Conv2D block ($c_5$) & (128, 96, 128) &DeConv2D block ($d_5$)& (256, 192, 64) \\
32&Conv2D block ($c_6$) &  (64, 48, 256)& DeConv2D block ($d_4$)& (128, 96, 128) \\
16&Conv2D block ($c_7$) & (32, 24, 512) &DeConv2D block ($d_3$)& (64, 48, 256) \\
8&Conv2D block   ($c_8$)& (16, 12, 1024)& DeConv2D block ($d_2$)& (32, 24, 512) \\
4&Conv2D block ($c_9$)& (8, 6, 2048) & DeConv2D block ($d_1$)& (16, 12, 1024) \\

\hline

\end{tabular}\label{network_arc}
\end{table}

 Here, we describe the key components of the architecture. The U-Net model consists of Conv2D blocks in the encoder and DeConv2D blocks in the decoder, operating at multiple resolutions, with skip connections linking corresponding layers between the two parts. A summary of the network architecture is provided in Table~\ref{network_arc}. The Conv2D block consists of following sequence of layers: {\ensuremath{\tt convolution}   $\rightarrow$ {\ensuremath{\tt Batch\,\,Normalisation } $\rightarrow$ {\ensuremath{\tt Activation} $\rightarrow$ {\ensuremath{\tt Dropout} $\rightarrow$ {\ensuremath{\tt convolution} $\rightarrow$ {\ensuremath{\tt Batch\,\,Normalisation } $\rightarrow$ {\ensuremath{\tt Activation}  $\rightarrow$ \ensuremath{\tt {Max-pooling}}.  The DeConv2D block consists of following sequence of layers: }{\ensuremath{\tt upsampling} $\rightarrow$ {\ensuremath{\tt concatenation} $\rightarrow$ {\ensuremath{\tt convolution} $\rightarrow$ {\ensuremath{\tt Batch\,\,Normalisation } $\rightarrow$ {\ensuremath{\tt Activation} $\rightarrow$ {\ensuremath{\tt Dropout} $\rightarrow$ {\ensuremath{\tt convolution} $\rightarrow$ {\ensuremath{\tt Batch\,\,Normalisation} $\rightarrow$ {\ensuremath{\tt Activation}.  At each Conv2D/DeConv2D block PReLU activation \citep{prelu:2015} is applied. Each Conv2D block downgrades the map size by half through {\ensuremath{\tt {Max-pooling}} operation. Similarly, each DeConv2D block doubles the size of the input maps of the previous layer. To transfer the information from the contracting path to the expansion path, a \textit{skip connection} is employed between hidden layers of encoder and decoder blocks at each resolution. For instance, $c_9$ is concatenated to the corresponding same-resolution maps at layer $d_1$. The same is followed between $c_{n}$th layer and $d_{10-n}$th layer. We use a kernel size of $4\times4$ for all convolutional layers, with a stride of 2 to control the downsampling and upsampling operations. The number of filters applied to each convolution is displayed in the third quantity of the corresponding output layer shape listed in Table~\ref{network_arc}. To prevent over-fitting, we use {\ensuremath{\tt Dropout}  that randomly zeroes some of the units with a probability of 10\%. We include {\ensuremath{\tt Batch \,\, Normalisation} in each hidden layer to ensure inputs of the next layer remain normalised, facilitating a stable and efficient training. Padding is applied to keep the same image size after the application of a convolution. We use the \textit{least absolute deviation} as the loss function to minimise the error between predicted values and true output values,
\begin{equation}
    \mathcal{L} = mean( |Y - f_{w,b}(\textbf{X})|),
\end{equation}
and Adam \citep{Adam:2014} optimiser is used in training to minimise the loss function. 

Since the needlet coefficient maps have different \Nside\ resolutions across the bands, we apply slightly modified versions of the network for each band while keeping the core network structure consistent. For instance, Table~\ref{network_arc} details the network design for needlet coefficient maps at \Nside\ = 1024. Bands 9 to 11,  use this full architecture, as input and output maps at those bands are at \Nside\ = 1024. For bands 6 to 8, where the input and output maps have a lower resolution of \Nside = 512, the network is adapted to start at layer $c_{2}$ and ends at layer $d_8$, producing a final output shape of (2048, 1536, 1), which corresponds to an output map resolution of \Nside = 512. This approach is similarly applied to other bands, adjusting the network architecture according to the respective \Nside\ resolution of the input and output maps for each band.

\section{Data}\label{data}
\subsection{simulations}\label{sims}
The simulated data sets used in network training and testing include the instrument specifications of the \planck\ CMB experiment.  The maps used in this work include the CMB and foreground contribution from thermal dust, synchrotron, free-free, AME, line emission of CO, thermal and kinetic Sunyaev-Zeldovich (tsz and ksz), and CIB. We do not include any point sources since \cite{Petroff:2020} has reported inclusion of point sources causes network training to be insufficient. The treatment of point sources in the analysis of CMB recovered from real data is discussed in Section~\ref{planck_application}. The CMB maps are Gaussian random realisations of given lensed CMB $C_{\ell}$s computed using CAMB\footnote{https://camb.info}. To capture the full range of the $\Lambda$CDM parameter space, in CAMB, we use a set of six $\Lambda$CDM parameters sampled from a Gaussian distribution, based on the best-fit values and 2$\sigma$ standard deviations reported by the \planck\ 2018 results \citep{planck-VI:2020}.

For the Galactic foregrounds, we use two set of models of thermal dust, synchrotron, free-free, and AME available in the Python Sky Model (\pysm) \citep{pysm:2017}. 
The sets consist of the following configurations:\\
 \\$\tt{Configuration-1}$: \ensuremath{\tt d1s1f1a2}: The \ensuremath{\tt d1} thermal dust model follows a single modified blackbody (MBB) spectrum that uses  545 \GHz\ dust maps for intensity, and temperature and spectral index maps from Commander analysis of \planck\ 2015 data \citep{planck-X:2016}. The synchrotron model \ensuremath{\tt s1} uses power law scaling. The model uses Haslam 408 MHz maps as synchrotron template \citep{Haslam_MR:2015} and a spatially varying synchrotron spectral index map derived by fitting 408 MHz and WMAP 23 GHz polarisation data using model 4 of \citep{Miville-Desch:2008}. The free-free model \ensuremath{\tt f1} uses the free-free emission template of Commander at 30 \GHz\ and scales it to other frequencies using a power law index of --2.14. The AME model \ensuremath{\tt a2} uses two spinning dust components of Commander \citep{planck-X:2016}, and frequency scaling is done using the SpDust2 code \citep{Ali:2009}.  \\
 \\$\tt{Configuration-2}$: \ensuremath{\tt d7s1f1a2}:   We replace the dust model of previous configuration with \ensuremath{\tt d7} that uses dust model as described in \cite{Draine:2016} based on different dust grains composition. This configuration is only used in Section~\ref{sec:imact_of_dust} as a test sample to assess the ability of the trained network to subtract the unseen dust complexity. We do not use this configuration in network training.\\ 

The foreground emissions are too complex in real-world scenarios, especially with complicated special variations of spectral parameters and amplitudes. We aim to train the network with the ability to predict the unseen data, which requires a sufficient generalisation of foreground training samples. 
In order to increase the independent training samples, we randomise the spectral parameters in each realisation, which matches within the prior knowledge of observational errors \citep{planck-XI:2014,planck-xxv:2016,mfiwidesurvey}. 
This helps the training process reduce dependency on any specific foreground template. We add maps of CO, tsz, ksz and CIB to the above configurations simulated using \pysm3, and due to the lack of scope to randomise these templates within the current simulation framework, we use only one realisation of these components. The impact of this is illustrated in detail in Section~\ref{reults}. We use \ensuremath{\tt co1} model for CO line intensities at central frequencies 115.3, 230.5 and 345.8 \GHz. The CO emission maps obtained from \planck\ data using Modified ILC  (\ensuremath{\tt MILC}, \cite{Hurier:2013}) method are used as templates. The CIB and SZ maps are simulated using \textit{WebSky} \ensuremath{\tt cib1}, \ensuremath{\tt tsz1} and \ensuremath{\tt ksz1} models \cite{Stein:2020}.  
We use band pass integration at each frequency band using \planck\ bandpass profiles at \planck\ frequencies.

We use 300 full focal plane simulations (FFP10) of \planck\ noise realisations \citep{planck-X:2016}.
All these noise maps are smoothed to the common beam FWHM and downgraded to \Nside = 1024 on \healpix\ grid \citep{Gorski:2005}. 

\subsection{Planck data}\label{sec:PR3}
We use all Low Frequency Instrument (LFI) maps at 30-70 \GHz\ and High Frequency Instrument (HFI) maps at 100--353 \GHz\ of \planck\ PR3 intensity data \citep{planck-I:2018} to apply to the trained ML model. All the maps are smoothed to a common beam FWHM=7.27$^{'}$ and downgraded to \Nside\ = 1024 before application to the ML model. Special care is taken when deconvolving the beam for LFI maps due to their larger beam sizes compared to the common beam; details of this procedure are discussed in Section~\ref{sec:LFIHFI}. We also use CMB temperature maps produced by Spectral Matching Independent Component Analysis (SMICA) and NILC provided in PR3 data release \citep{planck-IV:2020} for comparison with the CMB map produced in this work. When estimating the power spectrum from simulated maps, we employ the Galactic mask (\commask\ hereafter) from the \planck\ 2018 data release, which covers approximately 70\% of the sky. In the real data analysis presented in Section~\ref{planck_application}, we use the common intensity mask from \planck\ 2018 result. This mask is produced from CMB solutions of four pipelines (NILC, SMICA, COMMANDER and SEVEM, \citep{planck-IV:2020}), setting the thresholds of 3 $\mu K$ to standard deviation maps at large angular scales (FWHM = 80$^{'}$) and 10 $\mu K$ at small angular scales (FWHM = 10$^{'}$). We multiply point source masks of \planck\ \citep{planck-xxvi:2016} to a common intensity mask to produce a final confidence mask. This final mask excludes the Galactic plane and compact objects and retains approximately 76 \% of the sky. We refer to this mask \commaskk\ hereafter.

\section{data processing}\label{preprocessing}
We simulate a set of 1200 mock data in Section~\ref{sims} at \Nside\ = 1024. We use 1000 of these realisations for training, 100 realisations for validation and 100 realisations as test sets. All maps are decomposed into a set of needlet coefficient maps following Section~\ref{nilc}. Our network is designed to perform on 2D array. Therefore, we approximate the \healpix\ arrays of needlet coefficients to rectangular grids following the scheme of \cite{Wang:2022} before feeding the maps to the network. The \healpix\ maps in \nested\ ordering are arranged in 12 square grids of size \Nside $\times$ \Nside\ and then maps are placed in rectangular grids of size 4\Nside$\times$3\Nside. \cite{Wang:2022} has reported that this rearrangement of \healpix\ data can approximate the convolution over the sphere to a good extent.  The two-dimensional foreground cleaned CMB needlet coefficient maps recovered in the network output layer are projected back to the corresponding \Nside\ maps. A set of normalisation factors corresponding to different bands is applied to normalise the output CMB needlet coefficient maps to facilitate smoother training. The inverse transformation using the same factors is applied to network outputs to recover the CMB needlet coefficient maps during testing.  The output needlet coefficient maps are finally combined to recover the CMB map at \Nside = 1024 following Equation~\ref{EQ:NEEDLET-TO-CMB}. 

\section{Model training and CMB recovery}\label{reults}
Our network inputs are needlet coefficients of frequency maps comprising CMB, foregrounds, and instrument noise.  The expected outputs of the network are the needlet coefficients of the CMB temperature map at the common beam FWHM.   
We randomly select 1000 training samples in batches of 12 for network training. We randomly add FFP10 noise realisations to each of the sky realisations. We conduct experiments using two different sets of frequency combinations for training. Our first choice, referred to as $\tt Experiment-1$, includes four HFI frequency bands at 100, 143, 217, and 353 \GHz. The results obtained are discussed in this section. To further investigate the impact of incorporating both LFI and HFI frequency channels, we also perform $\tt Experiment-2$, which uses \planck\ frequency channels ranging from 30 \GHz\ to 353 \GHz. Details of this second experiment and results are discussed in Section~\ref{sec:LFIHFI}. 
The networks are trained for 1000 epochs\footnote{Each epoch refers to a single pass of the training dataset in batches of 12 through the network during the training procedure.}. The training procedure is performed in NVIDIA Dell R720 GPU and is completed in four days.  We set the learning rate to 0.1 at the start and gradually decrease to $10^{-6}$ at later epochs.  After completion, we find the training loss function reaches approximately below 0.5  for all the needlet bands. Validation loss is also to be found to reach to same level.

Once we validate the network with a validation set of 100 realisations, we apply the trained networks to test samples. We begin by examining the network trained with foreground configuration comprising  Galactic components only in the frequency range of 100--353 \GHz. We first compare all the results recovered from the network model with the corresponding results obtained from the NILC pipeline. In Figure~\ref{fig:input_recover_res}, we present the resulting foreground cleaned CMB map (upper left) and corresponding residual map (lower left) for one realisation of the test simulations. For comparison, in the right panel of Figure~\ref{fig:input_recover_res}, we present the NILC recovered CMB map (upper right) and corresponding residual (lower right) for the same realisation. Visually, the residual leakage in the ML-recovered CMB map is primarily concentrated at the Galactic plane, and is noticeably smaller than the residual leakage observed in the NILC result.
The power spectra of residual maps are more deeply assessed in the later part of this section.  
The mean absolute error (MAE) w.r.t the reference CMB maps, $< |Y^{Reference} - Y^{Predicted}|>$ is found to be 4.25 $\mu K$ for ML-recovered  CMB maps over \commask. The same is found to be 10.26 $\mu K$ for NILC-recovered CMB maps. 
We analyse the one-dimensional statistics by examining the probability density functions (PDFs) of the difference of needlet coefficients of ML-recovered CMB and reference CMB map at all needlet bands in Figure~\ref{fig:input_recover_dist}. 
 At band 1, we observe the largest discrepancy in two needlet coefficients, primarily due to the large-scale leakage of foregrounds to the recovered CMB. This effect arises from the approximation of \healpix\ maps to rectangular grids, which can distort large-scale foreground structures during convolution in the hidden layers. For the remaining bands, the dispersion of differences increases from larger (e.g., band 2) to smaller (e.g., band 11) angular scales. 

\begin{figure}
       \includegraphics[width=0.48\linewidth]{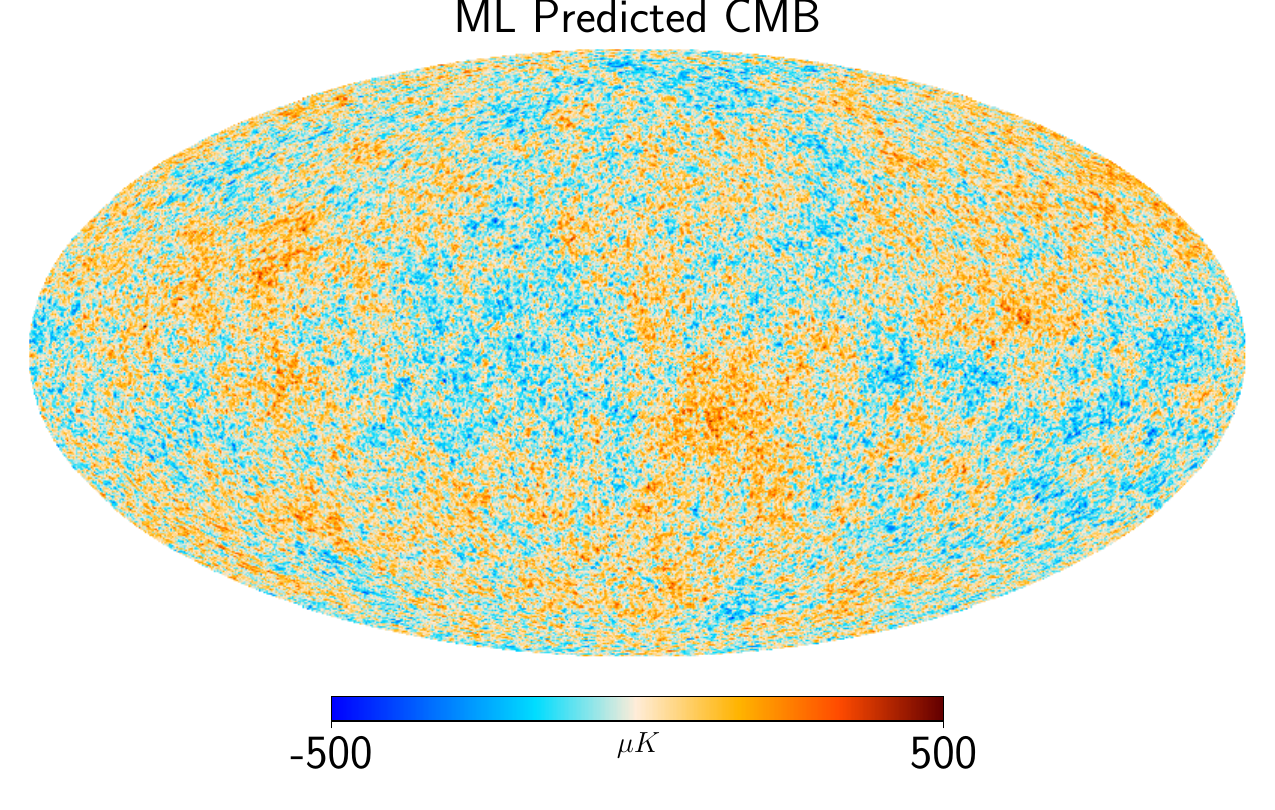}
       \includegraphics[width=0.48\linewidth]{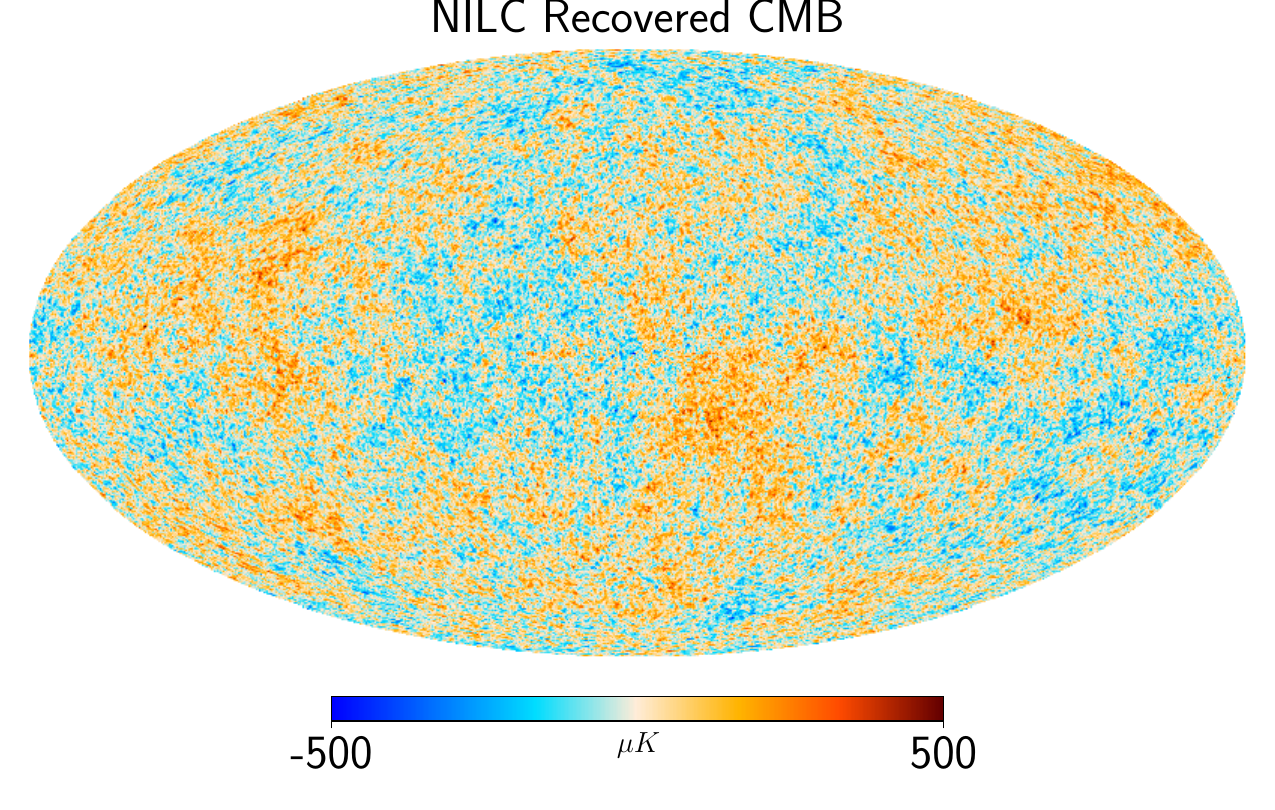}
       \includegraphics[width=0.48\linewidth]{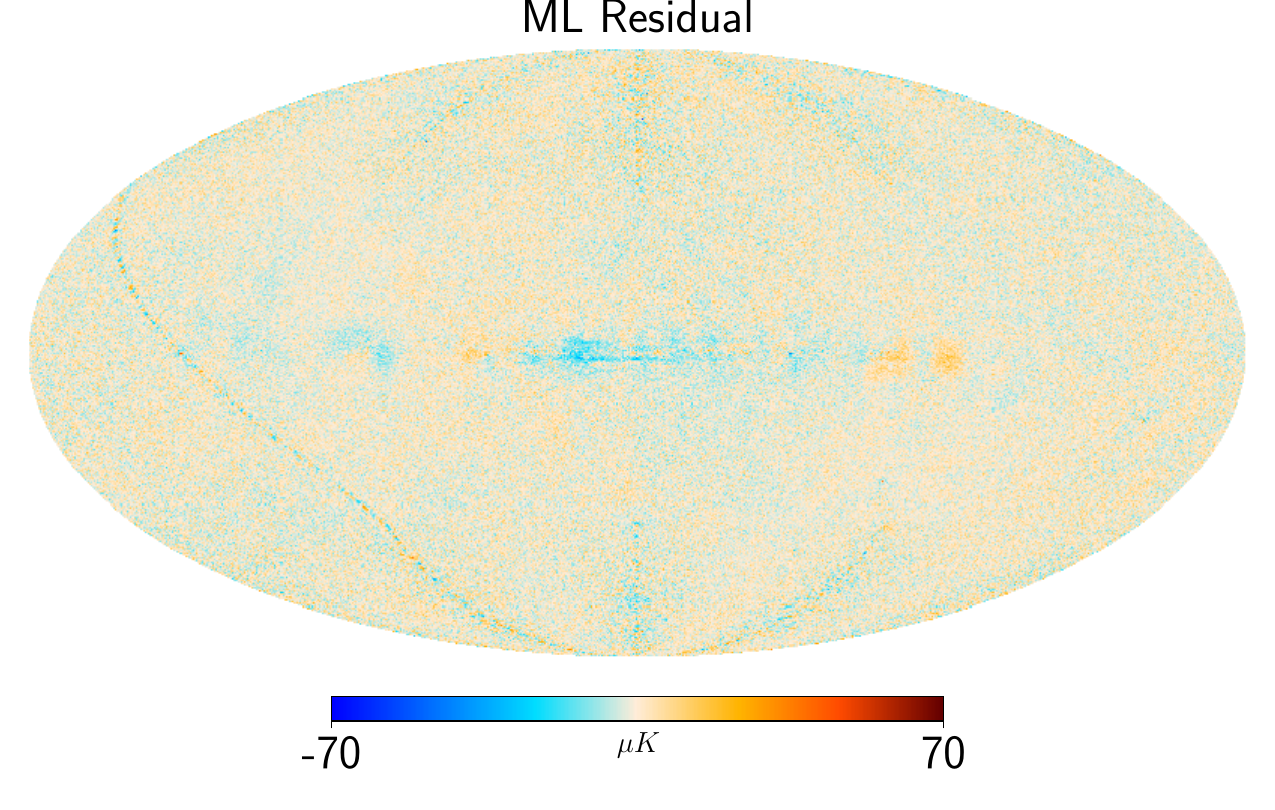}
       \includegraphics[width=0.48\linewidth]{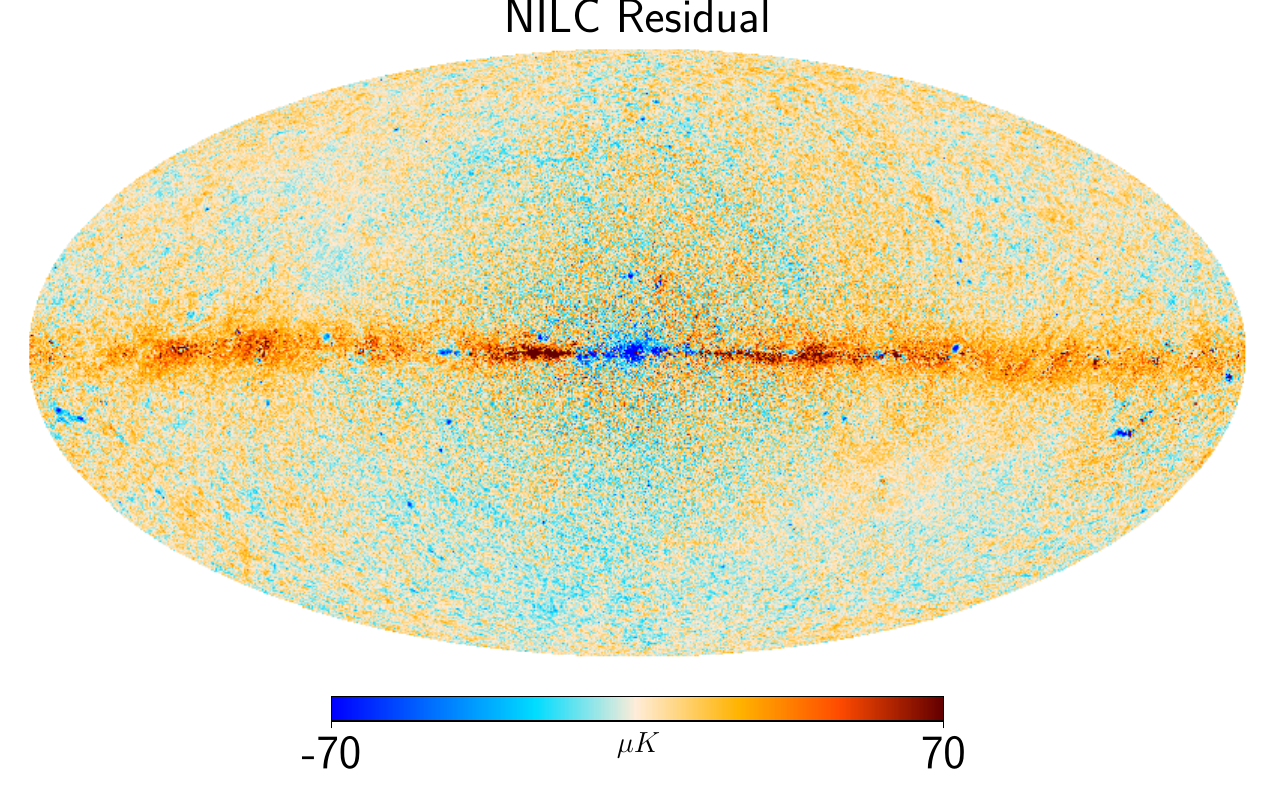}
\caption{The ML recovered CMB (upper left), NILC recovered CMB (upper right), residual leakage in recovered CMB for ML (lower left) and  NILC (lower right). }
\label{fig:input_recover_res}
\end{figure}
\begin{figure}
\includegraphics[width=\linewidth]{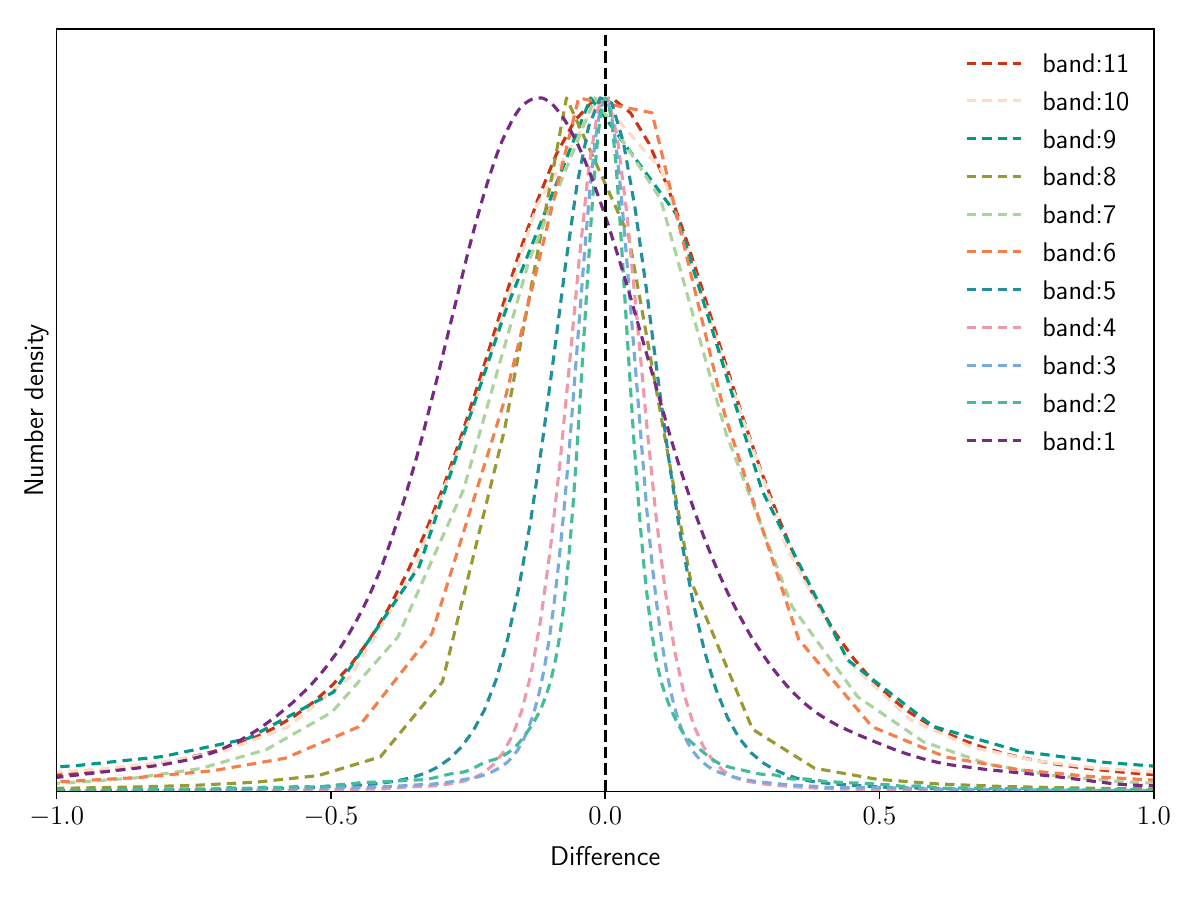}

\caption{The PDF of the difference between the needlet coefficient maps from network output and reference CMB at needlet bands for one test sample. The differences at bands 1 to 5 are divided by 10 to keep the x-axis range fixed for all bands for easy comparison. }
\label{fig:input_recover_dist}
\end{figure}

\begin{figure}
        \includegraphics[width=0.48\linewidth]{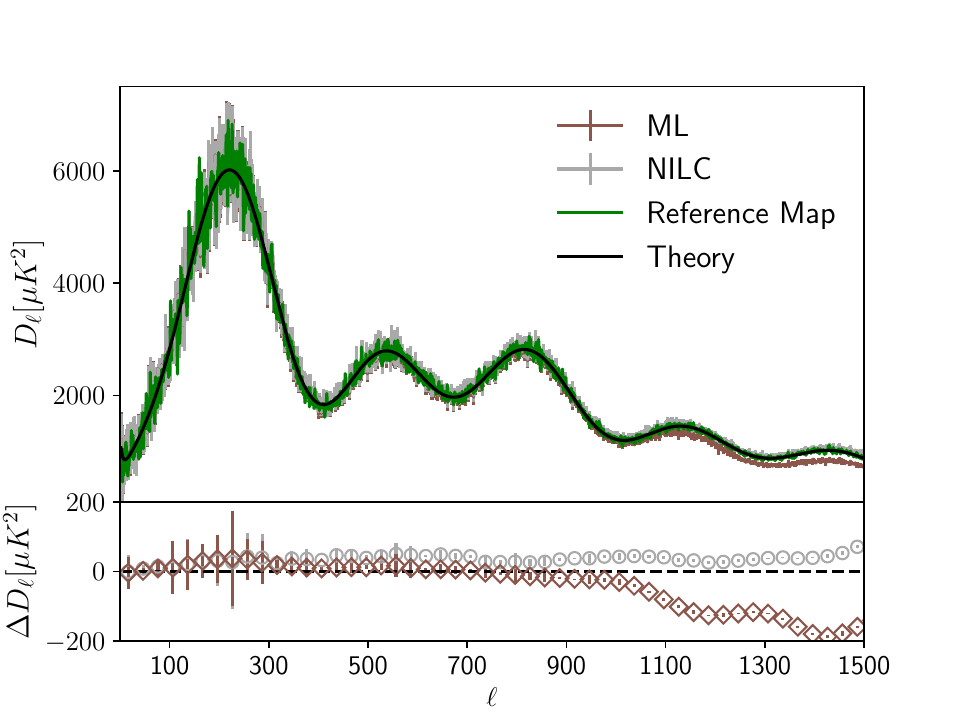}
       \includegraphics[width=0.48\linewidth]{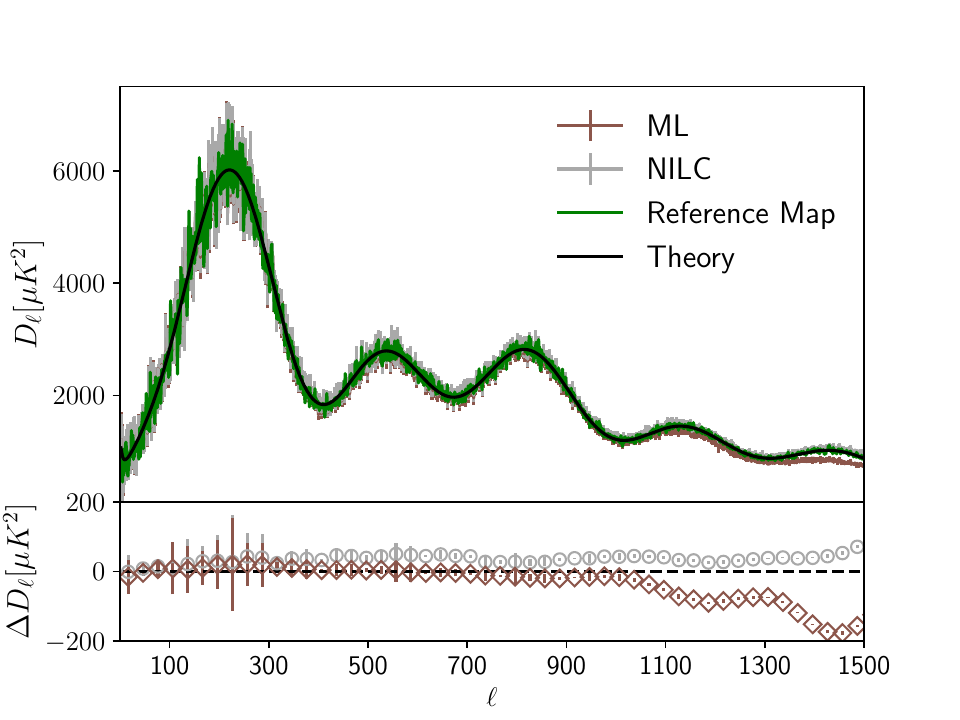}
       \includegraphics[width=0.48\linewidth]{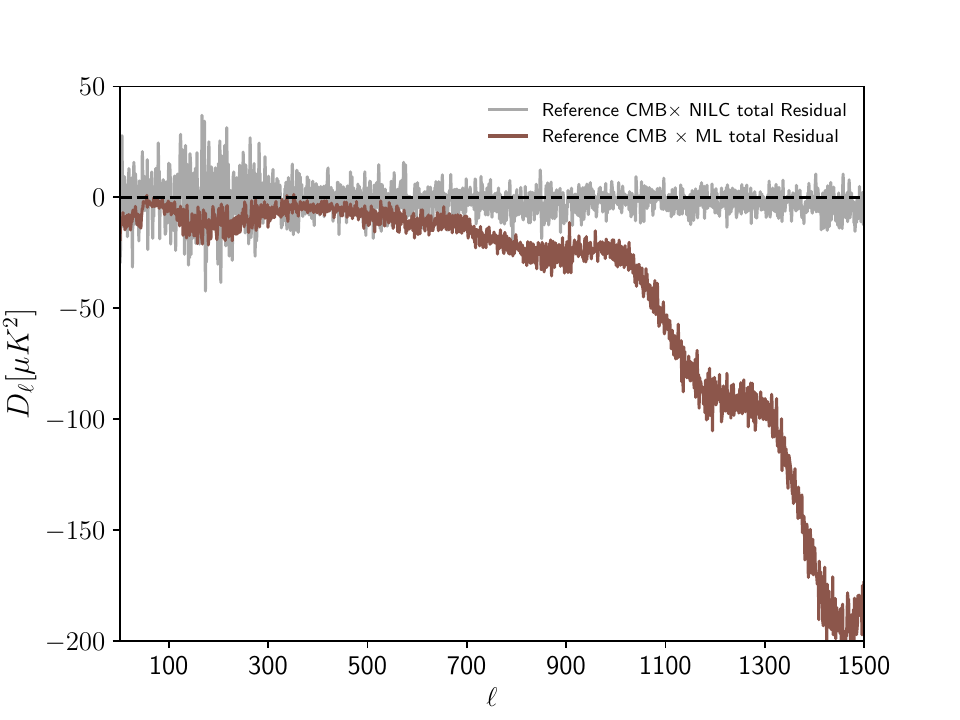}
       \includegraphics[width=0.48\linewidth]{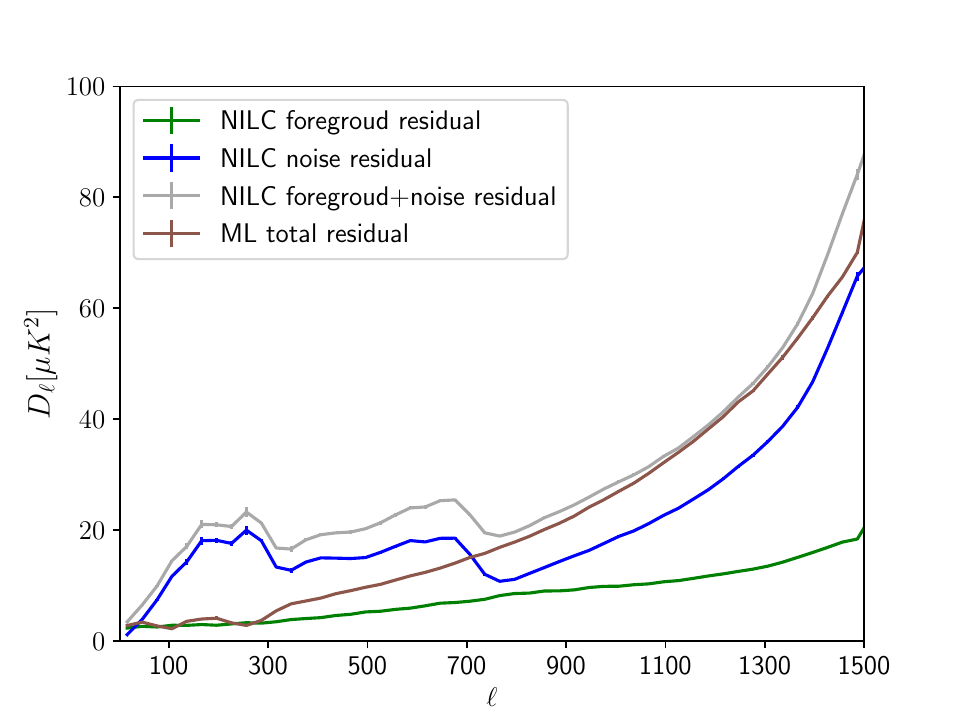}
\caption{ Upper left panel: Mean TT power spectra from 100 realisations of ML (brown ) and NILC (dark grey) recovered CMB maps and true reference CMB maps (green) estimated over \commask\ mask for foreground configuration comprising only Galactic components. In the bottom panel, we show the mean and 1$\sigma$ errorbars of deviation of recovered power spectra w.r.t power spectra of reference CMB maps binned with $\Delta \ell$ = 30.  The theoretical TT power spectrum is shown in a solid black line. Upper right panel: Same as upper left panel for foreground configuration comprising both Galactic and extragalactic components. Lower left panel: Bias in recovered CMB maps using both methods. The cross-correlation between the residual map of NILC and the reference CMB map (dark grey) is consistent with zero. A similar cross-correlation for the residual map for the network model (brown) is consistently negative, exhibiting leakage of the CMB signal to the residual at all scales. Lower right panel: Mean residual power spectra of noise (blue), foregrounds (green) and total (noise + foregrounds, in dark grey) present in NILC recovered CMB maps. The mean of the total residual power spectrum in ML recovered CMB maps is shown in brown. }%
\label{fig:input_recovered_TT}
\end{figure}

We evaluate the mean angular power spectrum (brown) from recovered CMB maps and compare it with the theoretical power spectrum (black) and mean power spectrum of the true reference CMB maps (green) in the upper left panel of Figure~\ref{fig:input_recovered_TT}. All power spectra are estimated using \xpol\ \citep{Xpol:2005} over 2$^{\circ}$ apodized \commask\ mask.  In the lower panel, we display the mean deviation and 1$\sigma$ errorbar of recovered power spectra w.r.t. mean power spectra of reference maps, binned with $\Delta\ell$ = 30. For comparison, we overplot the mean power spectrum (shown in dark grey) for recovered CMB maps using NILC. Across the full range of angular scales, the ML-recovered CMB maps exhibit a small multiplicative bias of less than 0.7\% in the mean angular power spectrum.  This bias arises primarily due to the use of the map-based loss function in network training. To investigate the bias in more detail, we cross-correlate the residual of the recovered map with the input reference map. The lower-left panel of Figure~\ref{fig:input_recovered_TT} shows the cross-power spectra for both the ML and NILC methods. While NILC residuals show no correlation with the reference CMB map, indicating negligible CMB signal leakage, the ML residuals exhibit a consistent negative correlation across all scales, indicating a biased recovery. We correct this bias using a quadratic fit: $<\Delta D_{\ell}> = \alpha <D^{Reference \,\,Map}_{\ell}> + \,\beta <D^{Reference\,\, Map}_{\ell}>^2$ to the residual power spectra as proposed in \cite{Petroff:2020}. This prescription significantly reduced bias at scales $\ell \lesssim$ 1100. A similar assessment at the map level may alter the information of the recovered CMB map, and therefore, we do not attempt to correct the bias at the map level. The same correction approach is applied consistently in Sections~\ref{sec:imact_of_dust}, \ref{sec:LFIHFI}, and \ref{planck_application} for simulated and real data respectively, using the correction parameters estimated in this section.
The bias-corrected power spectrum from ML predicted CMB maps agrees with the theoretical power spectrum at $\ell \lesssim 1100$. Below this scale, the network tends to underestimate the CMB power, as reflected by the systematically negative $\Delta D_\ell$ values.

Next, we assess the performance of the network trained with simulations that include extragalactic components. Compared to the previous configuration (which included only Galactic foregrounds), we observe a 10\% increase in MAE of the recovered CMB maps. In contrast, the corresponding increase in MAE for CMB maps recovered using NILC is only 5\%. Additionally, we find a larger bias ($<$ 1.3\%) in the power spectrum of the recovered CMB map as compared to the previous configuration. This increment is likely due to the limitation of using multiple realisations of extragalactic components in network training, which may restrict the model's ability to generalise well across all extragalactic features. In the upper right panel of Figure~\ref{fig:input_recovered_TT}, we present the mean power spectra for the ML-recovered CMB maps, the true reference CMB maps, and NILC-recovered CMB maps, along with the corresponding residuals in lower panel after applying the bias correction described earlier. Notably, the disagreement between the power spectrum from NILC results and ML results starts at lower $\ell (\sim 500)$ compared to the previous configuration. This earlier onset of divergence is again likely attributable to the limited variability of extragalactic components in the training data.  

We assess the power spectra of residual leakage to the recovered CMB map. The results are displayed in the lower right panel of Figure~\ref{fig:input_recovered_TT}. We show the mean power spectra of residuals from foregrounds (green), noise(blue), and foregrounds + noise (dark brown) for NILC and total residual (brown) for ML recovered CMB maps. 
The residual power spectrum for ML method is corrected for bias. 
The residual power spectrum for recovered CMB using the network model is less than the same for NILC total residual throughout the scales. After $\ell\sim1100$ this is mainly due to the less effectiveness of bias correction after $\ell~\sim$1100 for ML recovery discussed above. 
Below $\ell\sim700$, the residual power spectrum for the network model is even less than same of the foreground residual for NILC. This exhibits better performance of foreground subtraction of the network model that is based on the morphological features of non-Gaussian Galactic foregrounds that are dominantly present at large scales. On the contrary, the NILC method is blind to any spatial features of the foreground components.  However, CMB recovery of the network model is always prone to the spatial complexities of foregrounds. This has been discussed in the next section. %

\begin{figure}
 \includegraphics[width=0.33\linewidth]{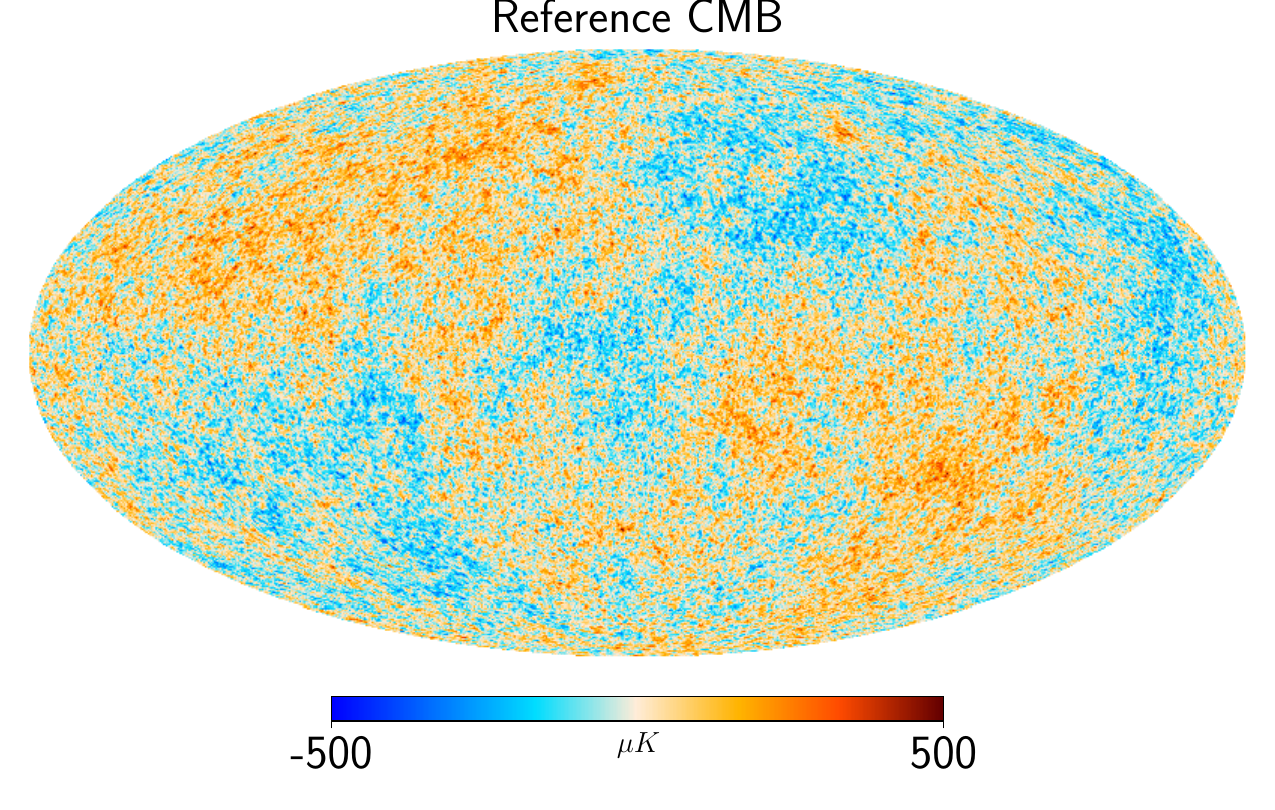}
 \includegraphics[width=0.33\linewidth]{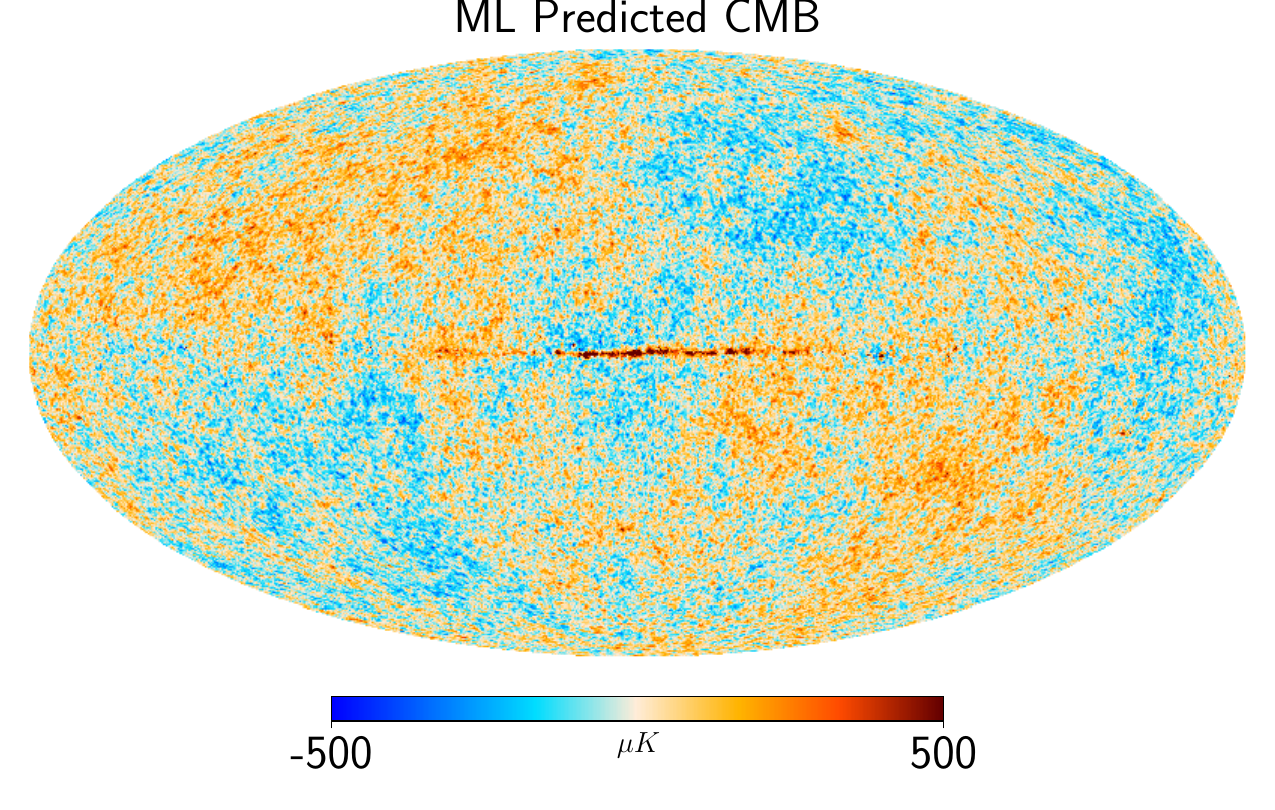}
 \includegraphics[width=0.33\linewidth]{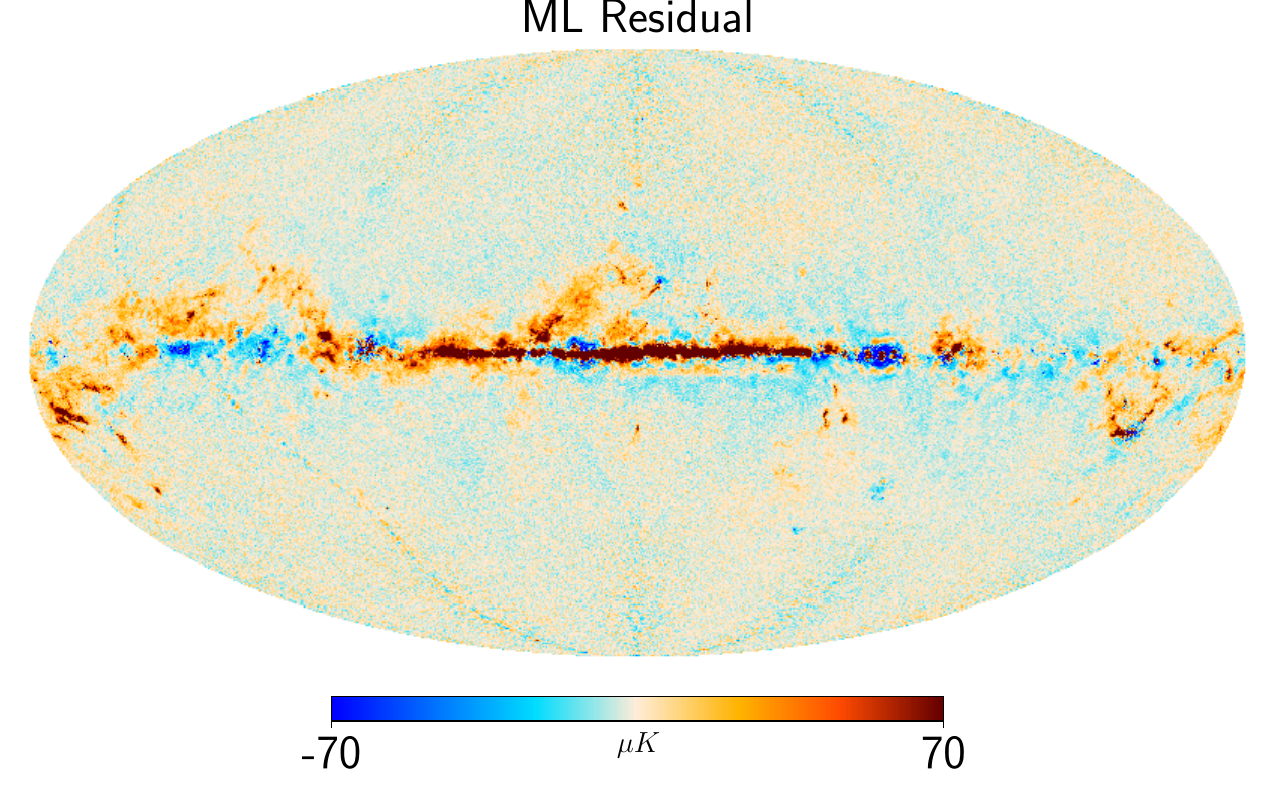}
 \caption{Left panel: Reference CMB map. Middle panel: ML recovered CMB map using $\tt{Configuration-2}$ as a test simulation. Right panel: Total residual in ML recovered CMB map.}
\label{fig:map_recover_d7}
\end{figure}

\begin{figure}[h]
 \includegraphics[width=\linewidth]{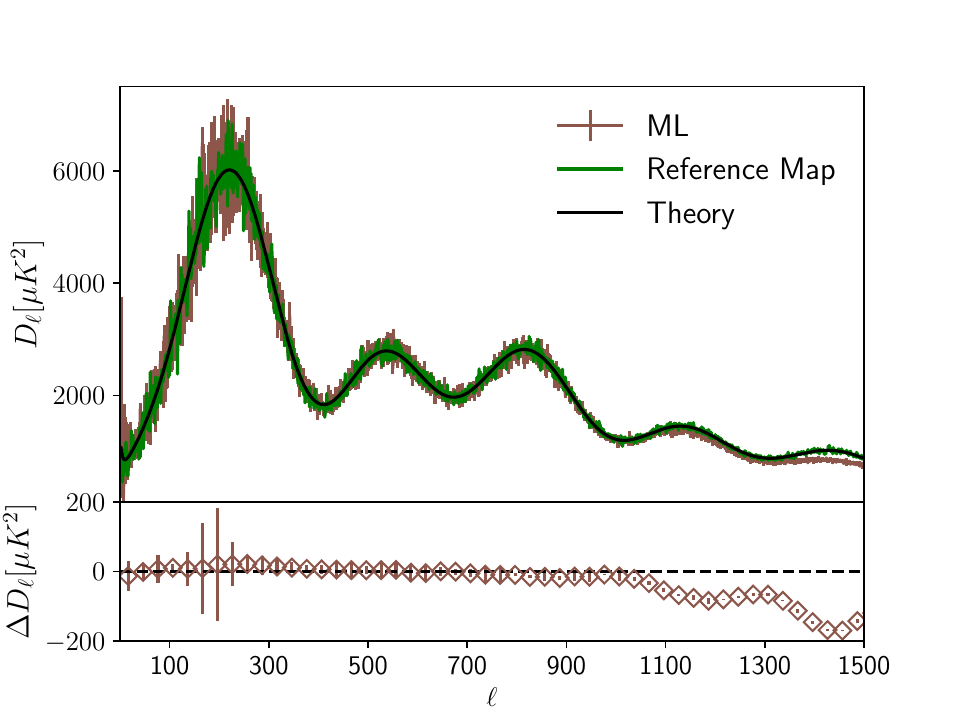}
\caption{The mean power spectrum of recovered CMB maps using $\tt{Configuration-2}$ as test simulations estimated over \commask\ mask. The green curve is the mean power spectrum of reference CMB maps. The lower panel shows the deviation of the recovered power spectrum from the true power spectrum of reference maps. }
\label{fig:spectrum_map_recover_d7}
\end{figure}

\begin{figure}[h]
 \includegraphics[width=0.5\textwidth]{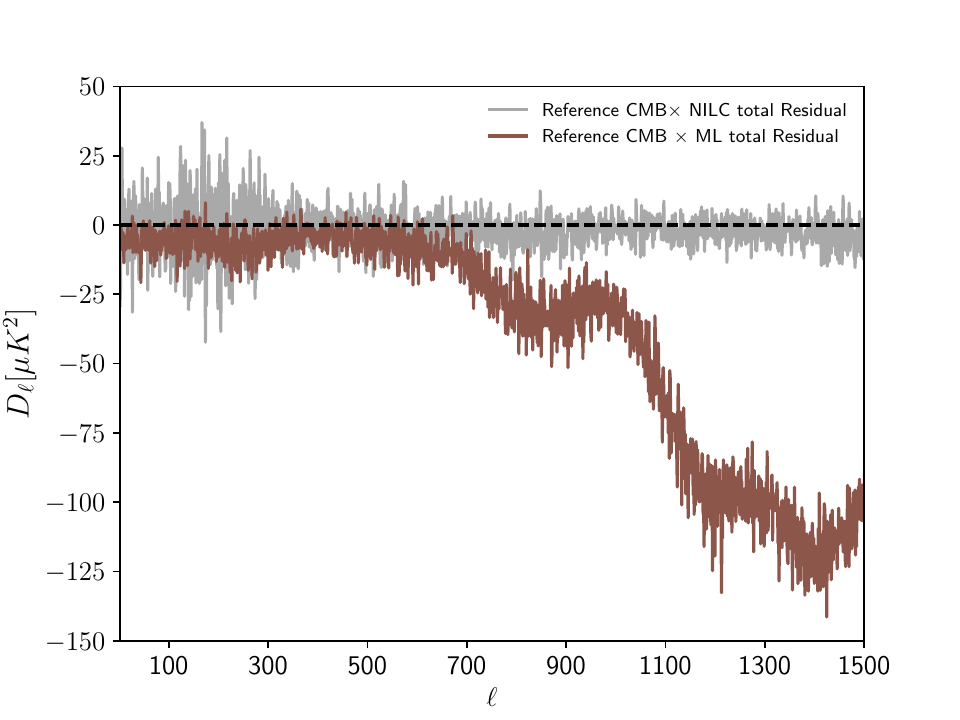}
 \includegraphics[width=0.5\textwidth]{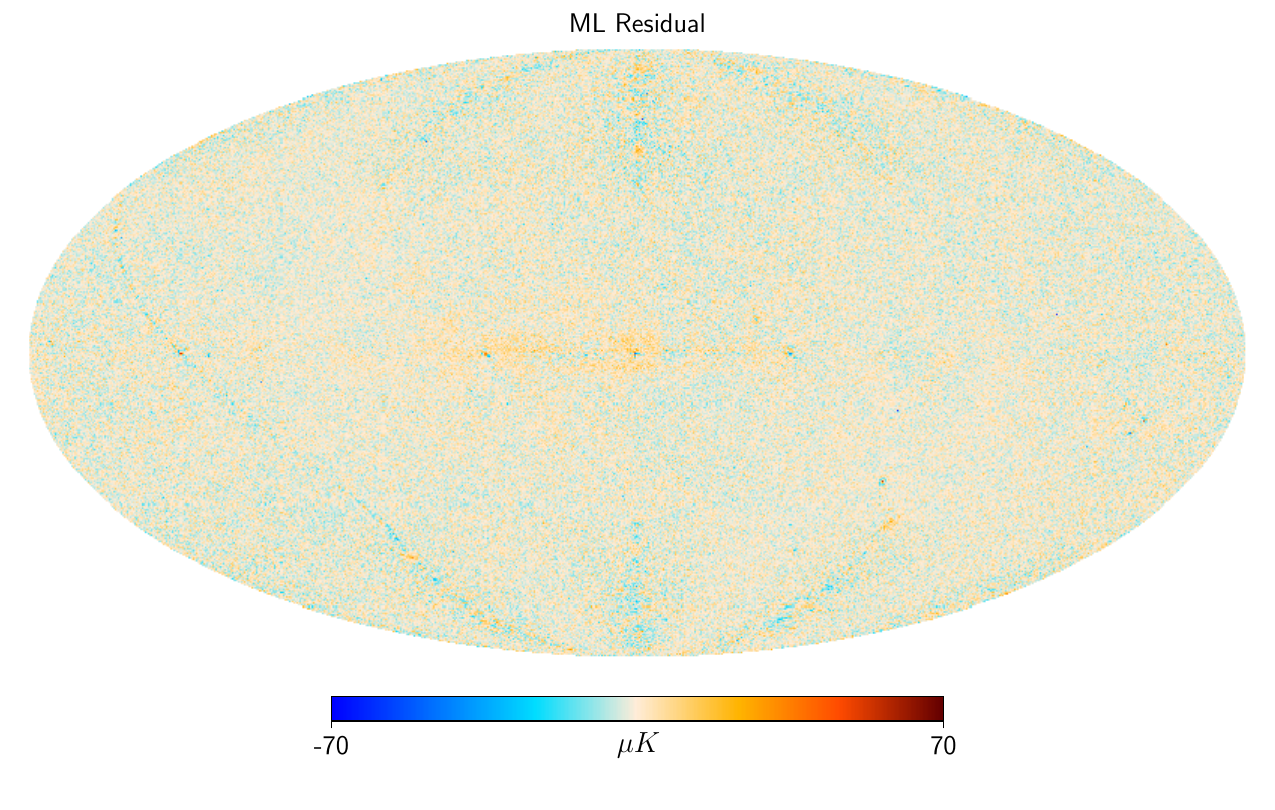}
\caption{Left panel: Bias in recovered CMB  maps for $\tt{experiment-2}$ using both cleaning method. The bias in the ML recovered CMB map is considerably reduced for $\tt{experiment-2}$ at $\ell< 1000$ as compared to $\tt{experiment-1}$. Right panel: The corresponding residual in the recovered CMB map.}
\label{fig:bias_residual_lfihfi}
\end{figure}
\begin{figure}[h]
 \includegraphics[width=\linewidth]{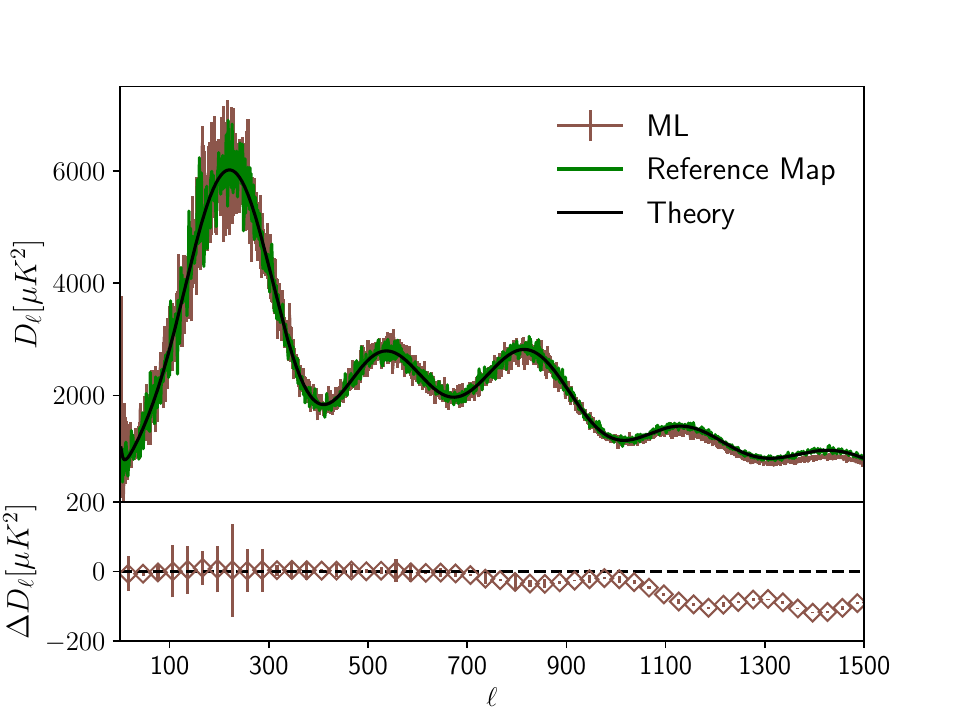}
\caption{Same as Figure~\ref{fig:spectrum_map_recover_d7} for predicted CMB using trained network for $\tt{experiment-2}$.}
\label{fig:spectrum_map_recover_d1_lfihfi}
\end{figure}

\subsection{Testing the impact of dust complexity}\label{sec:imact_of_dust}
The network described above was trained using $\tt{Configuration-1}$, which includes the \pysm\ \texttt{d1} dust model. Since thermal dust emission contributes significantly to foreground contamination in the 100–353 \GHz\ range, we test the robustness of the trained model by evaluating it under $\tt{Configuration-2}$, where the \pysm\ \texttt{d1} dust model is replaced with the more complex \texttt{d7} model. In Figure~\ref{fig:map_recover_d7}, we present the input reference map (left panel), the recovered CMB map (middle panel) and the residual (right panel) for one test sample. Compared to the previous result shown in Figure~\ref{fig:input_recover_res}, the residual is visibly larger, particularly in the Galactic plane region ($|b| < 15^{\circ}$).
The MAE increases to $\sim $14 $\mu K$ to which the largest contribution is seen from the first two needlet bands, most sensitive to the large-scale Galactic features. Despite this, the resulting power spectrum of recovered CMB shown in Figure~\ref{fig:spectrum_map_recover_d7} does not noticeably change with the updated dust model. The estimated power spectrum is corrected for bias following the same factors estimated in the previous section. This suggests that over 70\% intermediate and high latitude sky, the performance of the trained network remains robust, even when evaluated with a different dust model.


\subsection{Dependency on choice of frequency channels}\label{sec:LFIHFI}
In this section, we test whether additional information using more frequency channels improves the CMB recovery using the network model. Since \planck\ LFI  30, 44 \GHz\ and 70 \GHz\ channels provide additional information of foregrounds and CMB at angular scales $\ell \lesssim$ 650 and $\ell \lesssim$ 1000 respectively, we conduct $\tt{experiment-2}$, where we train the \texttt{Deep Needlet} model adding these three additional channels. We do not consider the last two HFI channels since noise power spectra are much larger than the CMB signal at 545 and 857 \GHz. For network training of $\tt{experiment-2}$, we follow a similar procedure discussed in Section~\ref{reults}. The LFI channels have a beam size that is larger than the common beam FWHM = 7.27$^{'}$.  As a result, convolving the LFI maps with a Gaussian of FWHM = 7.27$^{'}$ does not add any additional information at scales smaller than their respective beam sizes. Instead, this operation boosts the noise levels at small scales. To limit this effect, we selectively use only certain multipole ranges from the LFI channels when computing needlet coefficients, following the methodology of \citep{planck-IV:2020} (see Appendix B of their paper). Specifically, we retain only those multipoles for which the ratio of the LFI channel's beam transfer function to that of the common beam FWHM of 7.27$^{'}$ is less than 0.01. This cutoff of multipole range restricts the use of LFI channels across all needlet bands. Therefore, the LFI channels are used for those needlet bands where harmonic contents of the LFI maps span the entire bandwidth of the given needlet band. For instance, the 30 and 44 \GHz\ channels are used in up to needlet band 7. For a similar reason, 70 \GHz\ channel is used up to band 8. After validating the network with 100 validation sets, we evaluate the network using 100 test simulations. 
In Figure~\ref{fig:bias_residual_lfihfi}, we show the bias in the recovered CMB, represented by the cross-correlation power spectra between the recovered CMB and the reference CMB maps (left panel). The right panel shows the corresponding residual map for one test simulation. A reduction in residual leakage is observed, particularly at angular scales $\ell<700$, compared to $\tt{experiment-1}$ (see Figure~\ref{fig:input_recover_res}). The MAE is found to be 2.43 $\mu K$ that corresponds to $\sim40$\% reduction compared to the results in Section~\ref{reults}. 
We show the mean power spectrum from recovered maps and deviation w.r.t power spectra of reference CMB maps binned with $\Delta \ell = 30$ in Figure~\ref{fig:spectrum_map_recover_d1_lfihfi}. We see a small improvement in the recovered power spectrum w.r.t the power spectrum of the reference map. However, we do not see much improvement in small-scale bias at scales $\ell>1100$. This illustrates the incorporation of more information with additional channels little improves the network performance, and noise levels do not have much relevance in bias production. We discuss the role of instrument noise levels in network performance in Appendix~\ref{sec:App1} in more detail.

\begin{figure}
  \includegraphics[width=0.33\linewidth]{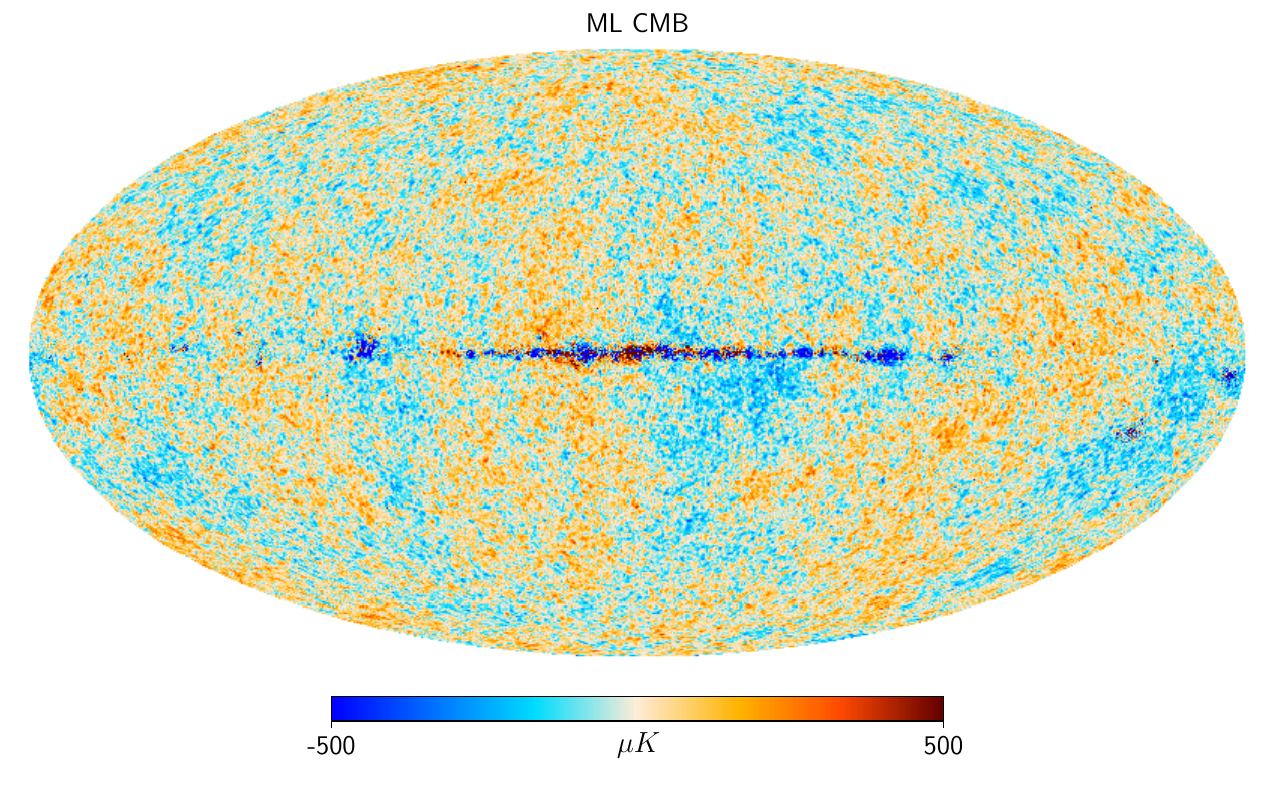}
 \includegraphics[width=0.33\linewidth]{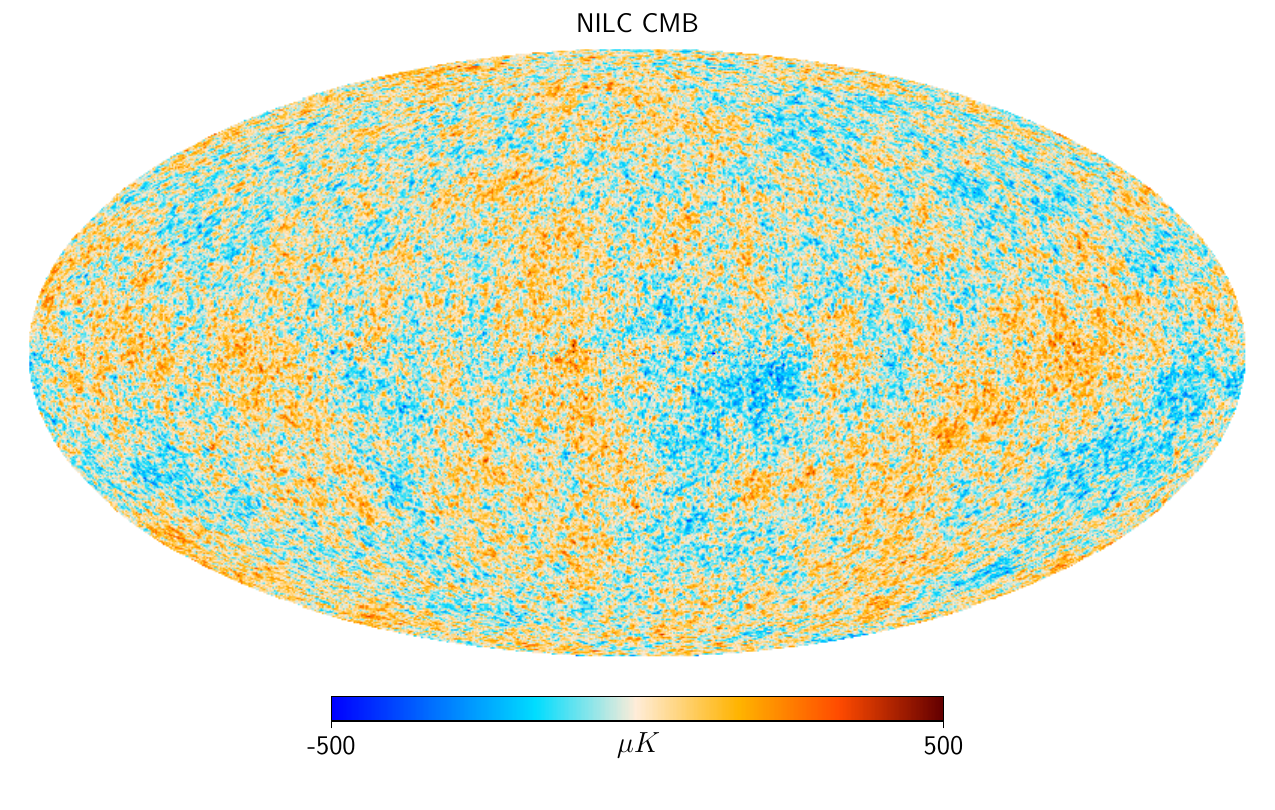}
 \includegraphics[width=0.33\linewidth]{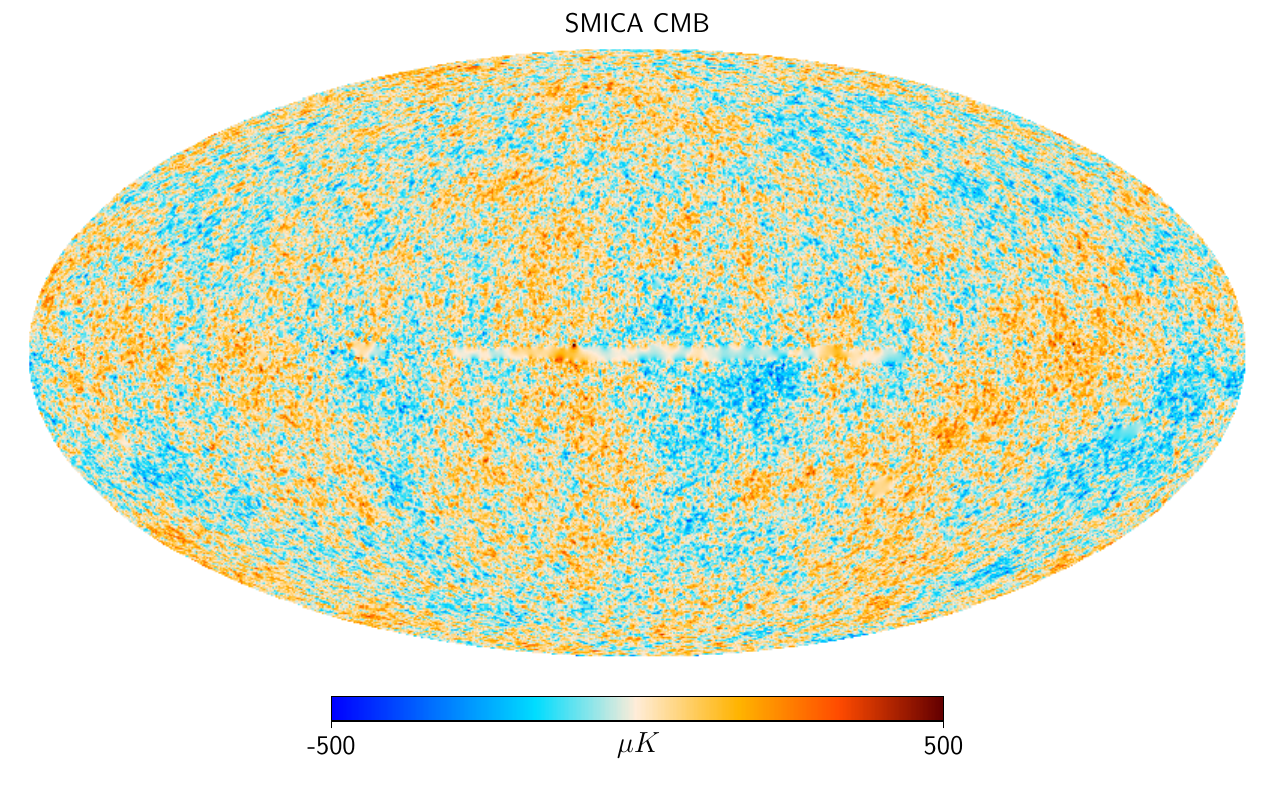}\\
 \includegraphics[width=0.33\linewidth]{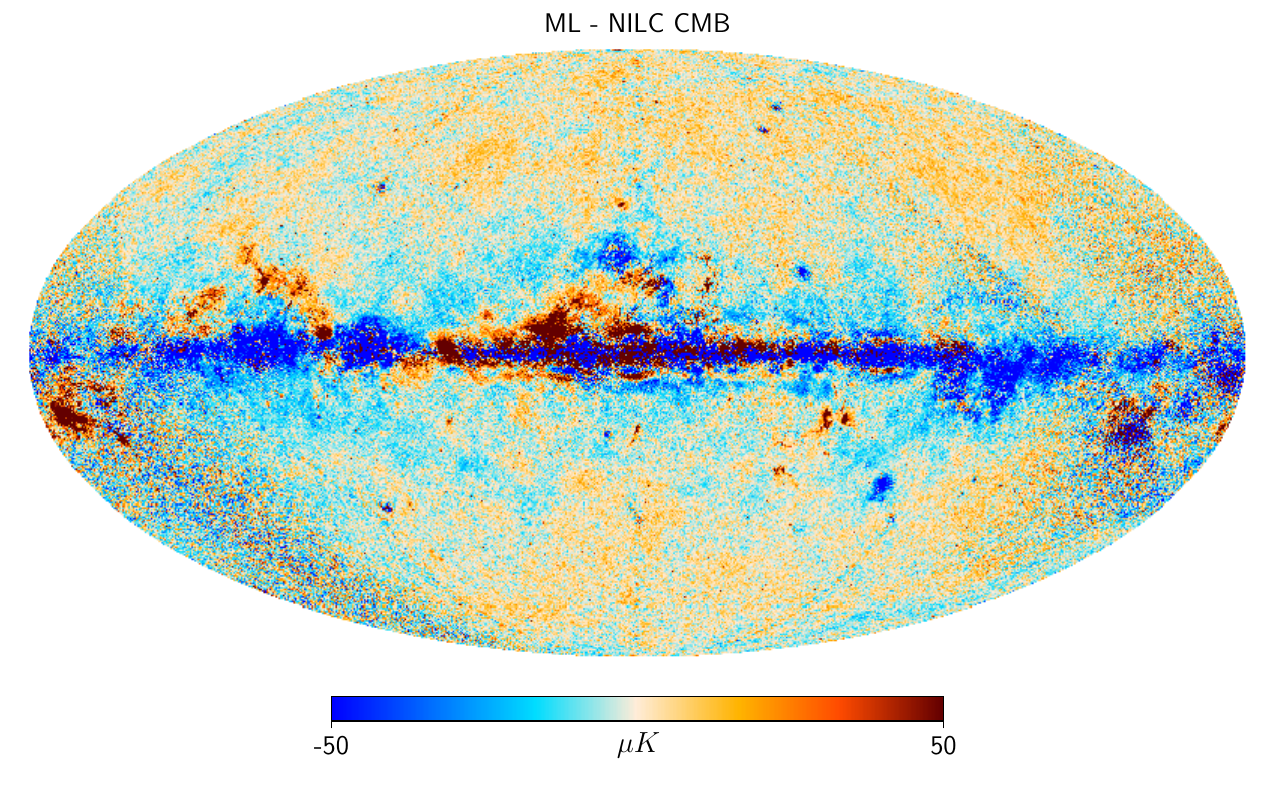}
  \includegraphics[width=0.33\linewidth]{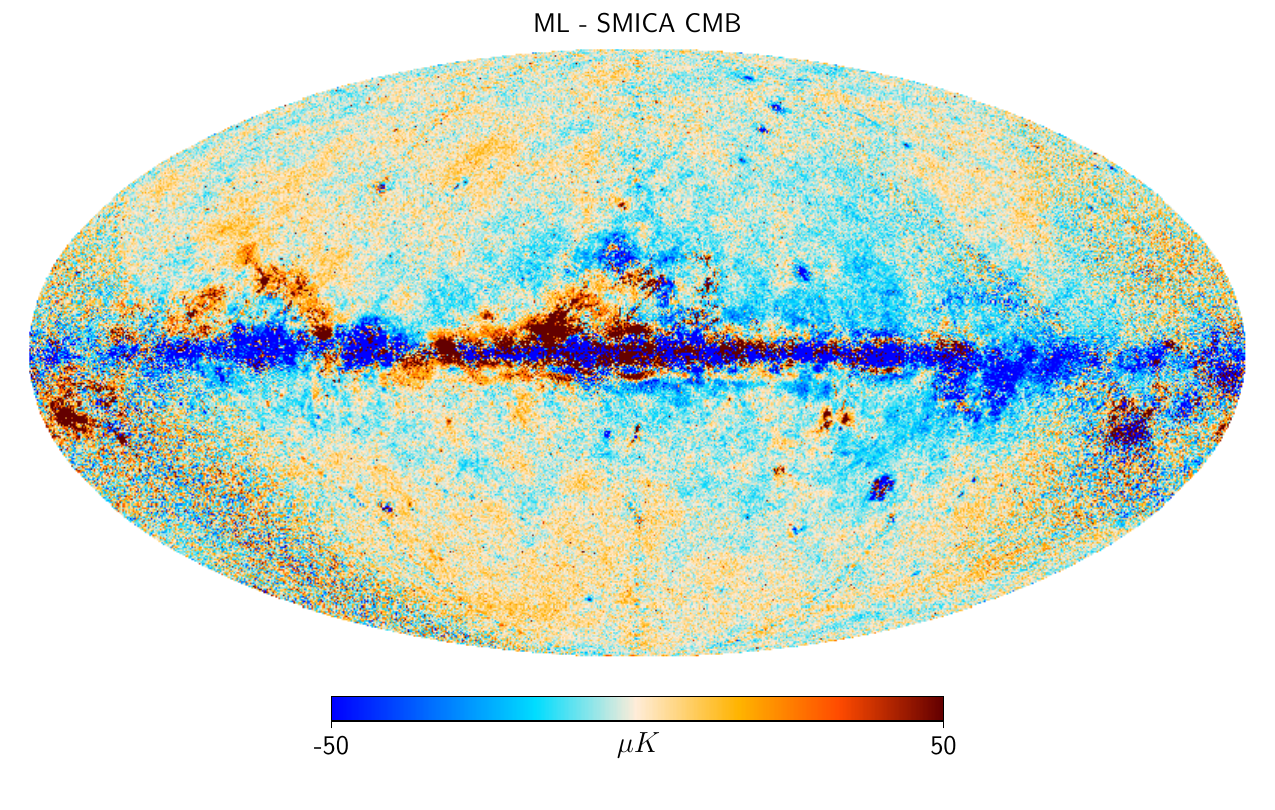}
 \includegraphics[width=0.33\linewidth]{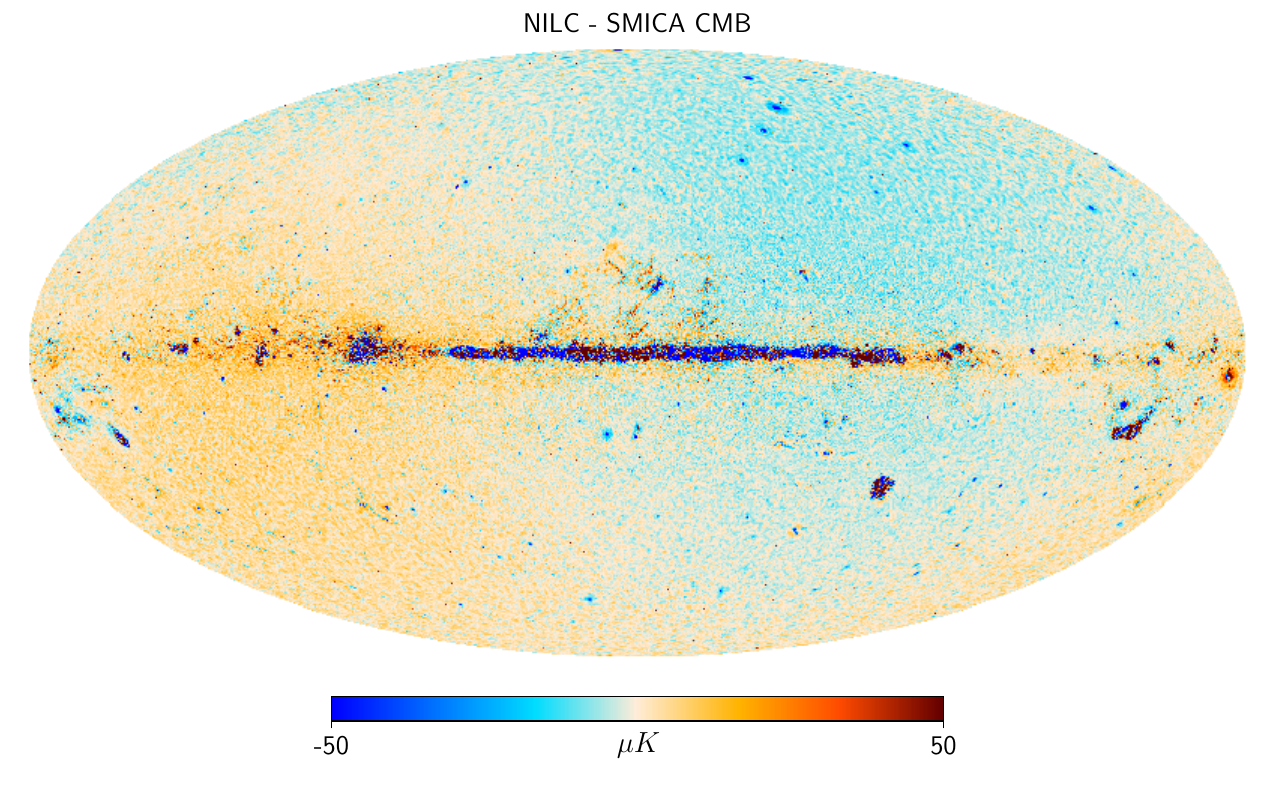} 
 \caption{Upper panel: ML recovered CMB from \planck\ data evaluated using network trained within 30-353 \GHz\ in Section~\ref{sec:LFIHFI}. The middle and right panel shows the official PR3 CMB map of \planck\ obtained from NILC and SMICA pipelines, respectively.  Lower panel: Pairwise differences between the recovered CMB of this work and CMB of \planck\ PR3 data obtained from NILC and SMICA pipelines. All the maps are smoothed at 10$^{'}$ resolution. }
\label{fig:map_recover_planck}
\end{figure}
\begin{figure}
       \includegraphics[width=\linewidth]{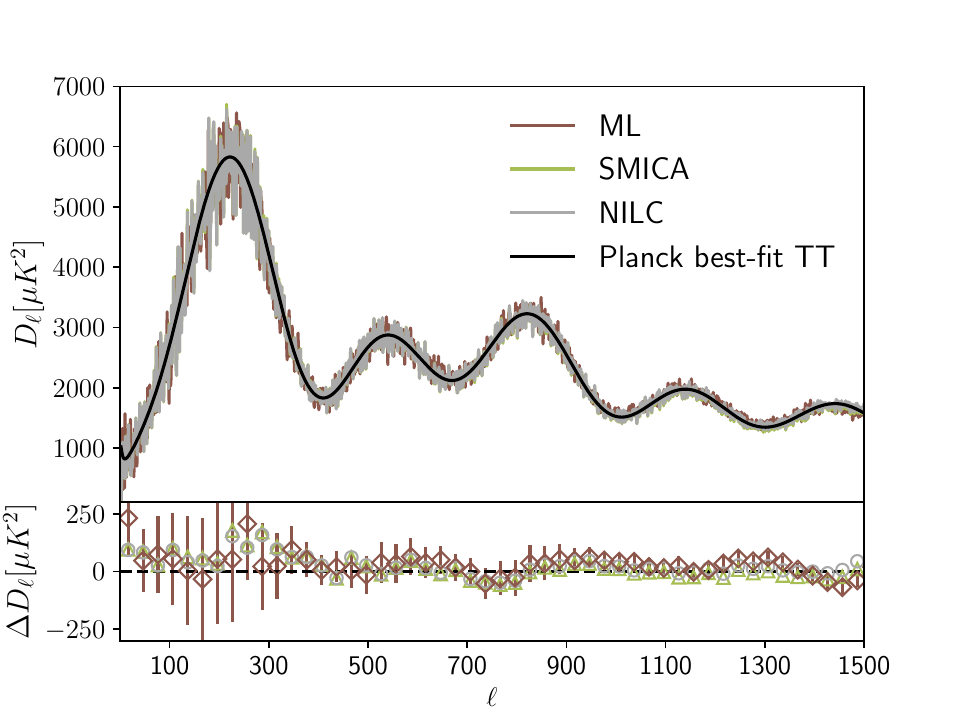}

\caption{The power spectrum from ML recovered CMB using \planck\ data between 30-353 \GHz\ estimated over \commaskk\ mask. In the lower panel, we present the deviation from \planck\ best-fit theoretical power spectrum within a bin of $\Delta \ell = 30$. The errorbars are cosmic variance at the corresponding bin. }
\label{fig:cltt_data}
\end{figure}
\section{Application to \planck\ data}\label{planck_application}
With the neural network model trained in Section~\ref{sec:LFIHFI}, we recover the CMB temperature map from \planck\ PR3 data. As illustrated in Section~\ref{sec:LFIHFI}, the network training considers the foreground model of $\tt{Configuration-1}$ and all LFI channels and HFI 100--353 \GHz\ maps. The final output map is at common beam FWHM = 7.27$^{'}$. We follow the same procedure discussed in Section~\ref{sec:LFIHFI} for LFI beam deconvolution. The resulting CMB map is displayed in the left panel of Figure~\ref{fig:map_recover_planck}. To compare with \planck\ legacy products of CMB temperature map, we show the NILC (middle panel) and SMICA (right panel) CMB maps of \planck\ 2018 results \citep{planck-IV:2020}. These maps are shown at 10$^{'}$ resolution. The lower panel shows all pairwise diﬀerence maps between each of the pipeline CMB maps. Comparison of the results from different methods is difficult since CMB maps of \planck\ results are inpainted near Galactic plane. Above the Galactic plane ($|b|>15^{\circ}$), the difference is present at small angular scales. 
The MAEs of ML-recovered CMB -- NILC CMB and ML-recovered CMB -- SMICA CMB are found to be 9.07 $\mu K $ and 8.67  $\mu K $ respectively after masking the Galactic pixels, which is close to the MAE of SMICA CMB -- NILC CMB that is 4.02 $\mu K $. 

We estimate the power spectrum of the reconstructed CMB map and compare it with SMICA and NILC CMB power spectra in Figure~\ref{fig:cltt_data}. All power spectra are estimated over \commaskk\ mask and corrected for respective beams. It is important to note that we do not consider the simulation of bright sources in training samples. Therefore recovery of the map at the pixels of compact objects is inaccurate. Therefore, we use \commaskk\ mask in the analysis of the recovered map that excludes pixels of the bright sources. The bias correction is implemented to estimate the power spectrum of ML recovered CMB following Section~\ref{sec:LFIHFI} for the sake of robustness. The best fit \planck\ power spectrum \citep{planck-VI:2020} is shown in a black solid line. In the lower panel, we show the deviations of power spectra from \planck\ best fit value binned with $\Delta \ell$ = 30. The 1$\sigma$ errorbar shown here is the cosmic variance limit. All power spectra match up to $\ell \simeq$ 900. The discrepancies between real data and simulated data used in training resulted in an increase presence of foreground residuals in the recovered CMB map. There is a small disagreement between the power spectrum of our recovered CMB and SMICA and NILC CMB maps at $\ell\gtrsim$ 900. To find the source of disagreement, we repeat the same analysis on \planck\ FFP10 simulations \citep{planck-III:2018}. There we find a small disagreement between the estimated power spectrum from the recovered map and \planck\ legacy data is present in large scales ($\ell <$ 300) instead of small scales. This illustrates that the source of disagreement for PR3 data might be caused by instrument systematics present in PR3 data, and therefore consistency check is beyond the scope of this paper since systematics are not taken into account in model training.   

\section{Discussion and Summary}\label{summary}
This work presents the development of a machine learning model in needlet space, referred to as \texttt{Deep\,\,Needlet}, designed to recover the CMB temperature map from multi-frequency observations in the millimetre and sub-millimetre bands.  The model is trained with simulated full sky maps that exhibit a diverse range of foreground spectral features and CMB temperature maps simulated using a spectrum of cosmological parameters sampled within 2$\sigma$ of the \planck\ 2018 best-fit values. We first evaluate the network on simulations that exhibit foreground complexity similar to the training set and compare the results with CMB maps recovered using NILC for the same realisations.  We find the MAE of the \texttt{Deep\,\,Needlet}–predicted CMB map is lower than that of the NILC recovered CMB map. Further insight into the ML method's effectiveness is gained by comparing the residual power spectra from both methods. At the multipole range $\ell \lesssim 1100$, the residual power spectrum from the machine learning model is lower than that of NILC.  The TT power spectrum of recovered CMB also shows good agreement with the true CMB spectrum up to multipole $\ell\sim 1100$. However, some systematic residuals remain at smaller angular scales. We also test the model's robustness by applying it to simulations with foreground complexity not included in the training set. In these cases, the network shows larger residuals near the Galactic plane, indicating reduced efficiency of the model in handling dust complexities that differ significantly from those seen during training. However, over approximately 70\% of the sky, outside the Galactic plane, the network is able to recover the CMB signal with adequate accuracy. This is supported by the power spectrum estimated over the \commask\ region, which continues to match the true CMB spectrum well up to $\ell \sim 1100$. We also explore the impact of incorporating additional large-scale information by including \planck\ LFI frequency channels in the network training. This extended frequency coverage leads to improved recovery of the CMB power spectrum, showing a better match with the reference CMB spectrum, particularly at large angular scales.

Our analysis highlights the inherent challenges of foreground feature extraction in machine learning–based models. Specifically, the use of a purely pixel-based loss function results in biased CMB recovery across all angular scales, with the bias becoming more pronounced at smaller angular scales. To investigate the origin of this bias, we analyse a model trained without instrument noise but with broader frequency coverage. The results suggest that the bias is intrinsic to the network architecture and training setup, largely due to reliance on pixel-wise loss optimisation. A promising direction to mitigate this issue is the adoption of a hybrid loss function that incorporates both pixel-level errors and statistical consistency, such as two-point correlation functions, as proposed in \cite{Makinen:2024}.

After being shown to work on simulations, we apply the network to \planck\ PR3 data. We find recovered CMB map is very similar to the CMB temperature maps from SMICA and NILC pipelines. The power spectrum of all CMB maps matches quite accurately at $\ell < 900$. After that scale, there is a small disagreement between the power spectra of recovered maps and NILC and SMICA CMB maps. This is due to a limitation of the use of instrument systematics in training sets that limits network performance in CMB recovery in the presence of instrument systematics in the real data. In this work, we only consider temperature maps to demonstrate the network performance. In future work, one avenue is to extend the presented technique to polarisation data. While the method performs well in several regimes, there are still limitations to be addressed, e.g.,  the presence of a known residual bias at smaller angular scales, and increased errors in regions with dust properties that differ from those seen during training, particularly near the Galactic plane. A potential solution is to incorporate a wider range of dust complexities into the training dataset. However, implementing this approach requires significant modifications to the training framework and is beyond the scope of this work; it is left for future investigation. 

\section{Acknowledgment}
All the computations in this paper are done using the High-performance computing (HPC) Nandadevi and Kamet (\url{https://www.imsc.res.in/computer_facilities}) at the Institute of Mathematical Sciences (IMSc), Chennai, India. Da acknowledges the financial support by the Department of Atomic Energy, India, that funded the postdoctoral position to the Department of Theoretical Physics at The Institute of Mathematical Sciences (IMSc). A financial support was provided by the Spanish Ministry of Science and Innovation under the project PID2020-120514GB-I00.
 This project uses the publicly available codes; \textit{cmbnncs} (\url{https://github.com/Guo-Jian-Wang/cmbNNCS}), PyTorch, PYSM, \healpix, CAMB, and Xpol. DA would like to thank Guo-Jian Wang and Benjamin Wandelt for useful discussion of ML techniques. DA would like to thank Jacques Delabrouille for his useful suggestions on NILC pipeline.


\appendix
\appendixpage
\section{Dependency on instrument noise}\label{sec:App1}
In this section, we test the impact of instrument noise on the performance of the ML model. \citep{Wang:2022} has discussed in details the performance of U-net architecture with 50\% and 70\% reduced instrument noise levels of \planck\ HFI channels (100--353 \GHz). They find the network performs with similar accuracy when they train the network using four HFI (100--353 \GHz) channels with \planck\ noise and using two HFI channels (143 and 217 \GHz) with 70 \% reduced \planck\ noise. On the contrary, we find the residual can only be reduced with the incorporation of more channels without reducing the \planck\ noise levels. It is important to note that the total residual in ML recovered CMB is a collective contribution of the biased prediction of CMB of ML model and foreground and noise present in the frequency maps, as discussed in the main text. In this section, we discuss whether the source of bias is \planck\ instrument noise or an intrinsic feature of the U-net architecture at small scales.  Therefore, we train the ML model without instrument noise and repeat our analysis of section~\ref{reults}.

\begin{figure}[h]
 \includegraphics[width=0.5\textwidth]{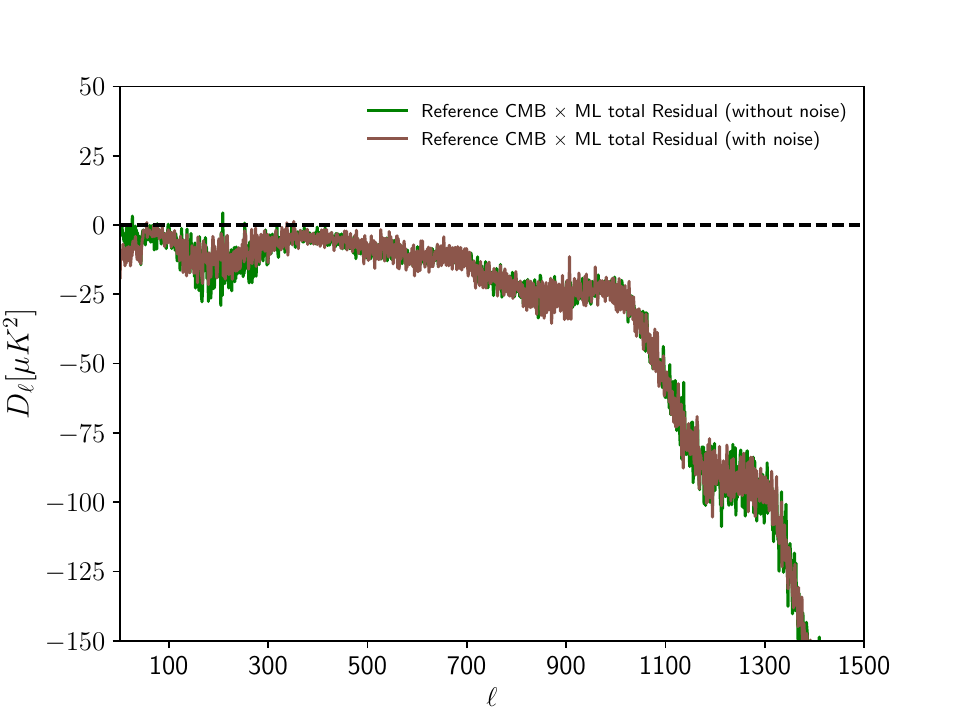}
 \includegraphics[width=0.5\textwidth]{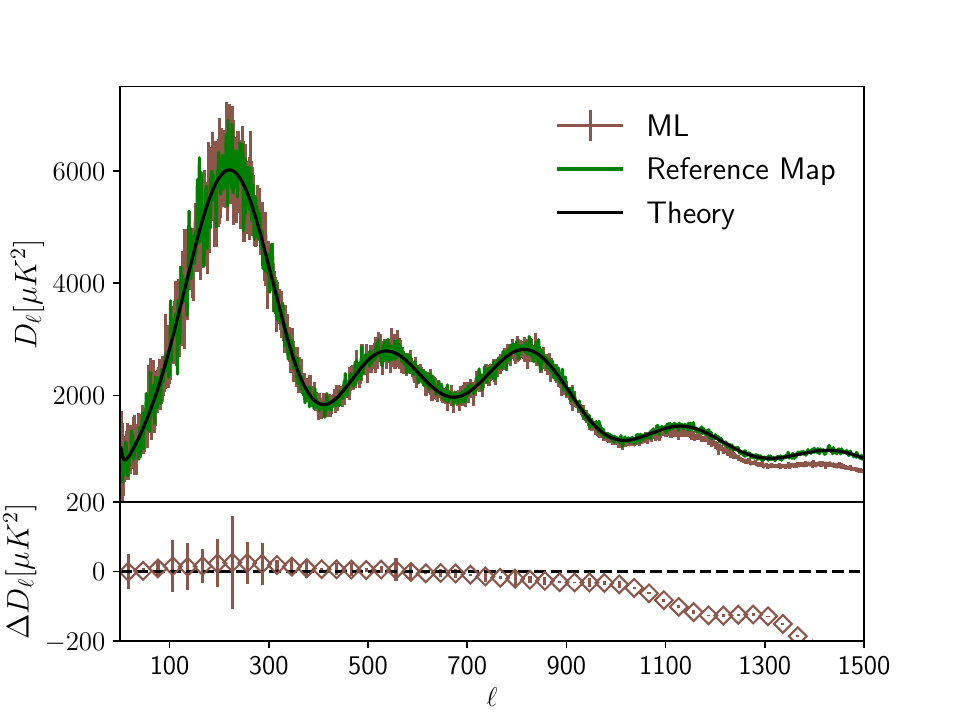}
\caption{Left panel: Comparison of bias in ML recovered CMB from trained network with (\textit{dark grey}) and without (\textit{green}) instrument noise. Right panel: Corresponding CMB power spectra and residual estimated over \commask\ and corrected for bias. }
\label{fig:spectrum_bias_no_noise_case}
\end{figure}
In the left panel of Figure~\ref{fig:spectrum_bias_no_noise_case}, we compare the bias in recovered CMB from the model trained with (dark grey) and without (green) instrument noise. We do not find any significant difference in cross-correlation between the input reference CMB and the residual maps for the two cases. In the right panel of Figure~\ref{fig:spectrum_bias_no_noise_case} we display the estimated power spectrum corrected for bias following the procedure in Section~\ref{reults}. Comparing this result with the upper right panel of Figure~\ref{fig:input_recovered_TT}, we see the estimated power spectrum matches with similar accuracy at $\ell < 1100$ for both cases, and approximately a similar bias is present in residual maps at small scales. However, we see a small improvement in residual bias at $\ell < 200$. This illustrates feature extraction at small scales using U-net architecture is intrinsic to the network model and relatively independent of \planck\ instrument noise levels.  This also hints that biased estimation of CMB map from the network model is a major contributor to residual leakage relative to \planck\ instrument noise. It is important to note that there is a subtle difference between our ML model and the model used in \cite{Wang:2022}. We use band-filtered signals in the training process, whereas \cite{Wang:2022} used full harmonic information in training. Since feature extraction depends on the type of training data sets used, two networks may cause different biases in recovered CMB maps. 

\bibliographystyle{JHEP}
\bibliography{reference}
\end{document}